\documentclass[tradiabstract,twocolumn]{aa} 
\usepackage{graphicx}
\usepackage{txfonts}
\usepackage{amsmath}
\usepackage{bm}
\usepackage{braket}
\usepackage{widetext}
\usepackage{fancybox}
\begin{document}
\thisfancyput(16cm,1.8cm){YITP-23-65}
\title{The gravitational force field of proto-pancakes}
\titlerunning{The gravitational force field of proto-pancakes}
\author{
Shohei Saga\inst{1,2,3,4}
\and
St{\'e}phane Colombi\inst{3}
\and
Atsushi Taruya\inst{5,6}
}
\authorrunning{S. Saga et al.}

\institute{
Institute for Advanced Research, Nagoya University, Furo-cho Chikusa-ku, Nagoya 464-8601, Japan
\and
Kobayashi-Maskawa Institute for the Origin of Particles and the
Universe, Nagoya University, Chikusa-ku, Nagoya, 464-8602, Japan
\and
Sorbonne Universit\'e, CNRS, UMR7095, Institut d'Astrophysique de Paris, 98bis boulevard Arago, F-75014 Paris, France
\and
Laboratoire Univers et Th{\'e}ories, Observatoire de Paris, Universit{\'e} PSL, Universit{\'e} de Paris, CNRS, F-92190 Meudon, France
\and
Center for Gravitational Physics and Quantum Information, Yukawa Institute for Theoretical Physics, Kyoto University, Kyoto 606-8502, Japan
\and
Kavli Institute for the Physics and Mathematics of the Universe (WPI), Todai institute
for Advanced Study, University of Tokyo, Kashiwa, Chiba 277-8568, Japan
}
\date{Received --; accepted --}
\abstract{It is well known that the first structures that form from small fluctuations in a self-gravitating, collisionless and initially smooth cold dark matter (CDM) fluid are pancakes. We study the gravitational force generated by such pancakes just after shell-crossing, and find a simple analytical formula for the force along the collapse direction, which can be applied to both the single- and multi-stream regimes. The formula is tested on the early growth of CDM protohaloes seeded by two or three crossed sine waves. Adopting the high-order Lagrangian perturbation theory (LPT) solution as a proxy for the dynamics, we confirm that our analytical prediction agrees well with the exact solution computed by direct resolution of the Poisson equation, as long as the caustic structure remains locally sufficiently one-dimensional. These results are further confirmed by comparisons of the LPT predictions performed this way to measurements in Vlasov simulations performed with the public code \texttt{ColDICE}. We also show that the component of the force orthogonal to the collapse direction preserves its single stream nature by not changing qualitatively before and after the collapse, allowing sufficiently high-order LPT acceleration to be used to approximate it accurately as long as the LPT series converges. As expected, solving Poisson equation on the density field generated with LPT displacement provides a more accurate force than the LPT acceleration itself, as a direct consequence of the faster convergence of the LPT series for the positions than for the accelerations. This may provide a clue on improving standard LPT predictions. Our investigations represent a very needed first step to study analytically gravitational dynamics in the multi-stream regime, by estimating, at leading order in time and space the proper backreaction on the gravitational field inside the pancakes.}
\keywords{gravitation - galaxies: kinematics and dynamics - dark matter}

\maketitle
\section{Introduction}
Cold dark matter (CDM)  is widely admitted to dominate the matter content of the Universe and is microscopically modelled as a self-gravitating collisionless fluid obeying the Vlasov-Poisson equations~\citep{1982ApJ...263L...1P,1984ApJ...277..470P,1984Natur.311..517B}.
Due to its initially virtually null local velocity dispersion, the CDM phase-space distribution function can be described as a three-dimensional sheet evolving in six-dimensional phase-space.  This sheet originally represents a single-stream flow, but as a consequence of the evolution under self-gravity, it can at some point self-intersect in configuration space. Such shell crossings mark the formation of singularities of various kinds, in particular pancake like structures accompanied by apparent divergences of the density field \citep[see, e.g.,][]{1970A&A.....5...84Z, 1982GApFD..20..111A, 1989RvMP...61..185S,1988PThPh..79..765G,1989ApJ...343...26M,2014MNRAS.437.3442H,2018JCAP...05..027F}. After the first shell crossings, the sheet repeatedly self-interacts and folds to form intricate multi-stream structures, that include filaments and dark matter haloes. Although numerical simulations have revealed a number of details on the dynamical history of dark matter, it is still difficult to develop an analytical theory capable of predicting the entire growth history of these structures in a fully self-consistent way, due to the highly nonlinear processes involved in multi-stream dynamics. As justified further below, an accurate description of the early stages of the evolution of multistream regions is fundamental to understand these processes and this must go through the calculation of the gravitational force field sourced by pancakes, which is the object of this article.

To approximate dark matter dynamics at large scale one usually relies on perturbation theory~\citep[hereafter, PT, see e.g.,][for a review]{2002PhR...367....1B}. PT has been widely used to predict, in the weakly nonlinear regime, large scale structure statistics such as the power-spectrum and higher order correlation. Many techniques have been developed in this framework~\citep[e.g.,][]{2006PhRvD..73f3520C,2007A&A...465..725V,2008PhRvD..77b3533C,2008ApJ...674..617T,2008PhRvD..77f3530M,2008PhRvD..78j3521B,2008JCAP...10..036P,2009PhRvD..80l3503T,2011PhRvD..83h3518M,2012PhRvD..85l3519B,2014PhRvD..89b3502B,2012PhRvD..86j3528T,2013PhRvD..87h3522V} and have been applied to observational data in order to constrain cosmological models~\citep[e.g.,][]{2011MNRAS.415.2876B,2017MNRAS.466.2242B,2019MNRAS.482.3497Z,2020JCAP...05..042I,2020A&A...633L..10T}.  In standard PT, a single stream flow is imposed, and the small parameter is the Eulerian density contrast. 
However, the single-stream approximation is valid only during the early phases of structure formation and its relevance can be questioned in the PT formalism~\citep[see, e.g.][]{2014PhRvD..89b3502B,2014JCAP...01..010B,2016PhLB..762..247N,2020MNRAS.499.1769H}. Beyond the single-stream approximation, it is challenging to incorporate backreactions from the multi-stream regions into the analytical predictions in generic situations~\citep[see e.g.,][for a recent review]{Rampf:2021rqu}. A way to account for multi-streaming on large-scale structure statistics consists in using effective field theory~\citep{2012JCAP...07..051B,2012JHEP...09..082C,2014PhRvD..89d3521H,2015PhRvD..92l3007B}. While effective field theory can provide for practical application meaningful constraints on cosmological models from observational data~\citep[e.g.,][]{2020JCAP...05..042I,2020JCAP...05..005D}, it involves free parameters that need to be calibrated with $N$-body simulations. 
For more rigorous treatment beyond phenomenological approach, one solid approach would be to consider high order velocity moments of Vlasov equations accounting for multistreaming. Recently, \citet{2023PhRvD.107f3539G,2023PhRvD.107f3540G} succeeded to make the problem analytically tractable, facilitating greatly convergence of PT predictions.
However, the approach developed by these authors still depends on simplifying assumptions and there is still room for improvement. Incorporating accurately multi-stream effects into statistical predictions of large-scale structure is thus now considered an important piece for properly extracting cosmological information from observations.

While pancakes generally correspond to the first stages of multistream evolution, filaments and then dark matter haloes represent the next ones, when shell-crossing happens subsequently along transverse directions of motion. The importance of accounting for multi-stream flows has therefore also been recognised in the formation process of protohaloes that allegedly develop monolithically during an early violent relaxation phase~\citep{1967MNRAS.136..101L} and subsequently, in the CDM scenario, merge hierarchically to form larger haloes with a universal density profile~\citep{1996ApJ...462..563N,1997ApJ...490..493N}. Although numerical investigations have revealed important properties of protohaloes, e.g., their power-law density profile,  $\rho(r)\sim  r^{-\alpha}$, with a logarithmic slope $\alpha\approx1.5$~\citep{1991ApJ...382..377M,2005Natur.433..389D,2014ApJ...788...27I,2017MNRAS.471.4687A,2018MNRAS.473.4339O,2018PhRvD..97d1303D,2018PhRvD..98f3527D,2021A&A...647A..66C,2022arXiv220911237D,2022MNRAS.517L..46W,2023MNRAS.518.3509D}, there is no exact analytical theory accounting for multi-stream dynamics inside dark matter haloes, despite the multiple approaches to the problem, for instance, self-similarity \citep{1984ApJ...281....1F,1985ApJS...58...39B,1995MNRAS.276..679H,1997PhRvD..56.1863S,1998ApJS..118..267Y,2004ApJS..151..185Y,2010PhRvD..82j4045Z,2010PhRvD..82j4044Z,2013MNRAS.428..340A} or entropy maximisation \citep{1967MNRAS.136..101L,2010ApJ...722..851H,2013MNRAS.432.3161C,2013MNRAS.430..121P}. 

One way to push the dynamics beyond shell crossing while preserving total mass conservation consists in using a Lagrangian approach that follows the motion of matter elements as functions of their initial position. Lagrangian PT (LPT), where the displacement field is the small parameter, has been widely employed to accurately describe the large-scale matter distribution in the quasi-linear regime, even beyond shell-crossing~\citep[e.g.,][]{1970A&A.....5...84Z,1989RvMP...61..185S,1992ApJ...394L...5B,1992MNRAS.254..729B,1993MNRAS.264..375B,1995A&A...296..575B,1994ApJ...427...51B}. Unfortunately, LPT becomes quickly inaccurate after shell-crossing because it does not account correctly of the force feedback inside multistream regions. Recently some progress has been achieved in this regard in the one-dimensional (1D) case, which corresponds to the simplest dynamical set-up, that is the pure pancake that reduces to the interaction between infinite parallel planes. In 1D, linear LPT is exact prior to shell-crossing~\citep{1969JETP...30..512N}, but subsequent multistream evolution still does not have a fully general analytical solution. However, it is possible to derive some approximate solutions asymptotically exact just after shell-crossing, based on LPT but extended beyond collapse and taking into account correctly the backreaction of the gravitational force inside the multistream region~\citep{2015MNRAS.446.2902C,2017MNRAS.470.4858T,2021MNRAS.505L..90R}. This article aims to generalize this post-collapse PT approach to three-dimension by calculating the gravitational force in protopancakes with the idea of developing a general analytical treatment of the early stages of multi-stream motion in the general case.

Contrary to the 1D case, one of the difficulties in developing a 3D post-collapse theory is the absence of an exact solution even before shell-crossing. This solution can be approached using sufficiently high order LPT, but as the system approaches the first shell crossing, the perturbative treatment worsens and its convergence speed decreases~\citep[e.g.,][for works on the convergence radius]{2014JFM...749..404Z,2015MNRAS.452.1421R,2021MNRAS.501L..71R,2022PhRvF...7j4610R}. Convergence speed depends strongly on the nature of initial conditions. It is facilitated when approaching quasi-1D initial conditions~\citep[e.g.,][]{2017MNRAS.471..671R,2018PhRvL.121x1302S} and is made more difficult when approaching axisymmetric configurations or spherical symmetry~\citep[e.g.,][]{2018PhRvL.121x1302S,2019MNRAS.484.5223R}.
In the simplified approach considered in the present work, we focus on pancakes seeded by a locally symmetric displacement field, a restrictive but still quite generic set up that applies for instance to high peaks of a Gaussian random field \cite[]{1986ApJ...304...15B}. To test our analytical predictions, we consider systems evolving from crossed sine waves initial conditions, that we already studied at collapse in \citet[][]{2018PhRvL.121x1302S} and slightly beyond shell-crossing in \citet[][hereafter, STC]{2022A&A...664A...3S}. For these systems, we found that LPT is able, at sufficiently high order, to provide an accurate description of the density distribution around shell-crossing so they represent a good ground for testing our analytical predictions for the gravitational force field. 

This paper is organised as follows. In Sect.~\ref{sec: basic}, we introduce the basic equations of motion in the Lagrangian description and the LPT framework. We also discuss important aspects of the gravitational force calculation. In Sect.~\ref{sec: formula Fx}, we provide a simple analytical formula for the force inside a pancake in the case the caustic structure is seeded by a locally symmetric displacement. In Sect.~\ref{sec: analytical ex}, our analytical predictions are tested in systems with sine-waves initial conditions, using high order LPT as a proxy for the dynamics. The approximation of the force field is compared to the exact solution of Poisson equation where the density field is generated by the LPT motion. This is followed in Sect.~\ref{sec: comparison} by comparisons of the analytical predictions to the force field directly measured in Vlasov-Poisson simulations performed with the public code {\tt ColDICE} \citep[][]{2016JCoPh.321..644S}.  Finally, Sect.~\ref{sec: summary} is devoted to the summary of important findings. To supplement the main text, Appendices~\ref{app: x_qx}, \ref{app: different G}, and \ref{app: force sim}, respectively test the validity of the series expansion at third order of the displacement field used in Sect.~\ref{sec: formula Fx}, the accuracy of our Green function approach used to compute the force in the theoretical calculations and finally the method we use to measure the force field in the {\tt ColDICE} simulations. 
\section{Basic equations and gravitational force}
\label{sec: basic}

We now introduce the basic equations in the Lagrangian description (Sect.~\ref{sec: setup}) and discuss important aspects of the gravitational force calculation in the framework of LPT (Sect.~\ref{sec: force}).

\subsection{Lagrangian dynamical framework}
\label{sec: setup}
We consider the Lagrangian equation of motion of a fluid element at the Eulerian comoving position $\bm{x}$ in the presence of gravity~\citep[e.g.,][]{1980lssu.book.....P}:
\begin{align}
\frac{{\rm d}^2\bm{x}}{{\rm d}t^2} + 2H \frac{{\rm d}\bm{x}}{{\rm d}t} &= -\frac{1}{a^{2}}\bm{\nabla}_{x} \phi(\bm{x}), 
\label{eq:basic_EoM}
\end{align}
where the quantities $a$, $H(t)=a^{-1} {\rm d}a/{\rm d}t$, and $\phi(\bm{x})$ are the scale factor of the Universe, the Hubble parameter, and the Newton gravitational potential, respectively, and the operator $\bm{\nabla}_{x} = \partial/\partial\bm{x}$ is the spatial gradient in Eulerian space.
The gravitational potential is related to the matter density contrast $\delta(\bm{x}) = \rho(\bm{x})/\bar{\rho}-1$ with $\bar{\rho}$ the background mass density, through the Poisson equation:
\begin{align}
\bm{\nabla}^{2}_{x}\phi(\bm{x}) &= 4\pi G a^{2}\,\bar{\rho} \delta(\bm{x}) . 
\label{eq:Poisson}
\end{align}
The Lagrangian description relates the initial, Lagrangian position, $\bm{q}$, of each mass element to the Eulerian position at time $t$, $\bm{x}(\bm{q},t)$, through the introduction of the displacement field, $\bm{\Psi}(\bm{q},t)$. In this framework, the Eulerian position $\bm{x}$ and velocity $\bm{v}$ are given by
\begin{align}
\bm{x}(\bm{q},t) &= \bm{q}+ \bm{\Psi}(\bm{q},t) , \label{eq: def Psi} \\
\bm{v} &= a\frac{{\rm d}\bm{\Psi}}{{\rm d}t} .  \label{eq: vel Psi}
\end{align}
Assuming homogeneous initial density, $\rho(\bm{q})/{\bar \rho}=1$, mass conservation reads ${\rm d}^3\bm{q}= \left( 1 + \delta(\bm{x})\right)\, {\rm d}^3\bm{x}$ before the first shell-crossing time  $t_{\rm sc}$. Hence, 
\begin{align}
1 + \delta(\bm{x}) =  \frac{1}{J} , \label{eq: delta-J}
\end{align}
where the quantity $J = \det{J_{ij}}$ is the Jacobian of the matrix $J_{ij}$ defined by
\begin{equation}
 J_{ij}(\bm{q}, t) =\frac{\partial x_{i}(\bm{q}, t)}{\partial q_{j}} = \delta_{ij} + \Psi_{i,j}(\bm{q}, t) .
 \label{eq: def Jij}
\end{equation}
The first occurrence of $J=0$ determines the first shell-crossing time $t_{\rm sc}$.

Until the first shell-crossing, we can employ a perturbative treatment to predict the fluid motion, namely Lagrangian perturbation theory~\citep[LPT, e.g.,][]{1970A&A.....5...84Z,1989RvMP...61..185S,1992ApJ...394L...5B,1992MNRAS.254..729B,1993MNRAS.264..375B,1995A&A...296..575B,1994ApJ...427...51B}. In LPT, the displacement field, $\bm{\Psi}$, is considered a small quantity, which is systematically expanded as
\begin{align}
\bm{\Psi}(\bm{q},t) = \sum_{n = 1}^{\infty}\bm{\Psi}^{(n)}(\bm{q},t) . 
\end{align}
Assuming that the fastest growing modes dominate, the perturbative solution is approximated quite well by the form~\citep[see, e.g.,][and references therein]{2002PhR...367....1B}:
\begin{align}
\bm{\Psi}^{(n)}(\bm{q},t) = D^{n}_{+}(t)\, \bm{\Psi}^{(n)}(\bm{q}) , \label{eq: Psi expansion}
\end{align}
with the time-dependent function $D_{+}(t)$ being the linear growth factor. Substituting Eq.~(\ref{eq: Psi expansion}) into Eq.~(\ref{eq: vel Psi}), the velocity field is given by
\begin{align}
\bm{v}(\bm{q},t) = a\, H\, f\,\sum_{n = 1}^{\infty}n\,D^{n}_{+}(t) \,\bm{\Psi}^{(n)}(\bm{q}) , \label{eq: vel LPT}
\end{align}
where we define the linear growth rate by $f(t) \equiv {\rm d}\ln{D_{+}}/{\rm d}\ln{a}$.

In equations (\ref{eq: Psi expansion}) and (\ref{eq: vel LPT}) the $n^{\rm th}$ order displacement $\bm{\Psi}^{(n)}$ is computed recursively by exploiting equations (\ref{eq:basic_EoM}) and (\ref{eq:Poisson}) using well known algebraic techniques \citep[see, e.g.][]{2012JCAP...12..004R,2014JFM...749..404Z,2015MNRAS.452.1421R,2015PhRvD..92b3534M}. Specific expressions for the three sine wave case examined below are given in STC and we do not consider it necessary to recall them.
The Lagrangian framework we are using in this article relies in practice on perturbation theory as a proxy of the dynamics. However the analytic expressions computed in Sect.~\ref{sec: formula Fx} are more general  in the sense that they apply (asymptotically, that is just beyond shell-crossing) to any non-degenerate displacement field locally symmetric and Taylor expandable up to third order in the Lagrangian position around the singularity of interest.
\subsection{Gravitational force}
\label{sec: force}
To facilitate resolution of Poisson equation (\ref{eq:Poisson}), we decompose it into two independent equations:
\begin{align}
\bm{\nabla}^{2}_{x}\phi_{\rm p} &= 4\pi G\bar{\rho}a^{2} \, (1+\delta(\bm{x})) , \label{eq: delta phi}\\
\bm{\nabla}^{2}_{x}\bar{\phi}   &= - 4\pi G\bar{\rho}a^{2} . \label{eq: bar phi}
\end{align}
Then, 
\begin{align}
\phi(\bm{x}) = \phi_{\rm p}(\bm{x}) + \bar{\phi}(\bm{x}) .
\end{align}
The first and second terms of the right-hand side of this equation are respectively the gravitational potential coming from the total density $\bar{\rho}(1+\delta(\bm{x}))$  and the negative background density $-\bar{\rho}$ as counter-term. After solving Eqs.~(\ref{eq: delta phi}) and (\ref{eq: bar phi}), the gravitational acceleration, also abusively referred to here as force $\bm{F}(\bm{x}) = - \bm{\nabla}_{x}\phi(\bm{x})$, is given by
\begin{align}%
\bm{F}(\bm{x})
&= 4\pi G a^{2}\bar{\rho} \left( \int\frac{{\rm d}^{d}\bm{x}'}{2^{d-1}\pi} \frac{(1+\delta(\bm{x}'))(\bm{x}-\bm{x}')}{|\bm{x}-\bm{x}'|^{n}} + \frac{1}{d}\bm{x} \right) \label{eq:neweul}
, \\
&= 4\pi G a^{2}\bar{\rho} \left( \int\frac{{\rm d}^{d}\bm{q}}{2^{d-1}\pi}\frac{\bm{x}-\bm{x}'(\bm{q})}{|\bm{x}-\bm{x}'(\bm{q})|^{n}}  + \frac{1}{d}\bm{x} \right)
, \label{eq: Newton}
\end{align}
where $d$ represents the dimension of space ($d=2$ or 3 considered in this paper). In the second line, we used mass conservation in $d$-dimensional space: $(1+\delta(\bm{x}'))\,{\rm d}^{d}\bm{x}' = {\rm d}^{d}\bm{q}$. The second term in parentheses represents the gravitational force arising from the background density in Eq.~(\ref{eq: bar phi}).

Thanks to the change of variable $\bm{x} \to \bm{q}$, the expression of the force in Eq.~(\ref{eq: Newton}) does not depend explicitly on the density contrast nor does it require a detailed knowledge of the multistream structure, which is, regardless of its complexity, implicitly contained in the function $\bm{x}(\bm{q})$. This considerably facilitates the numerical calculation of the integrals without the need to solve a multivalued problem. Note that this property was exploited before to compute the gravity field from caustic rings~\citep[see e.g.,][]{1998PhLB..432..139S,1999PhRvD..60f3501S,2006PhRvD..73b3510N,2007PhRvD..76b3505N,2008PhRvD..78f3508D,2009PhLB..675..279O,2018PhRvD..98j3009C,2012arXiv1205.1260T} and the calculations we perform here are analogous. 

While integral (\ref{eq: Newton}) seems simpler to estimate than integral (\ref{eq:neweul}), because it is performed in Lagrangian space, it remains challenging to compute it analytically and the main goal of the present work is to find explicit expressions approximating it.

Another way to estimate the gravitational acceleration consists in simply computing the second time derivative of the Lagrangian displacement (provided that it is sourced by pure gravity). Consider the presence of $n_{\rm S}$ streams at Eulerian position $\bm{x}$. In this case, the acceleration can be formally written as the local average of the time derivatives of velocities over all the streams weighted by the density of each stream:
\begin{align}
 \bm{F}(\bm{x}) =  \frac{\sum_{i=1}^{n_{\rm S}}\rho_{i}(\bm{x})\, \bm{\Gamma}_ i(\bm{x}) }{\sum_{i=1}^{n_{\rm S}}\rho_{i}(\bm{x})}
  , \label{eq: LPT acc multi}\\
  \bm{\Gamma}_ i(\bm{x}) \equiv \frac{{\rm d}(a\, \bm{v}_{i}(\bm{x}))}{{\rm d}t}, \label{eq: acc each flow}
\end{align}
where the quantities $\rho_{i}$ and $a\, \bm{v}_{i}$ stand for the density and peculiar velocity of $i^{\rm th}$ stream, respectively. If one has access to the exact solution of the dynamics, this expression is somewhat trivial since, in this case, $\bm{\Gamma}_ i(\bm{x})=\bm{\Gamma}_ j(\bm{x})$, $i \neq j$. On the other hand, if the accelerations in Eq.~(\ref{eq: LPT acc multi}) are given by second time derivatives of the LPT displacement computed at some order, the force in Eq.~(\ref{eq: LPT acc multi}) does not generally agree with Eq.~(\ref{eq: Newton}) applied to the same displacement field, even in the single stream regime. Indeed, the LPT solutions are derived not by directly solving the Poisson equation as in Eq.~(\ref{eq: Newton}) but by perturbatively solving the Lagrangian equations of motion.

Equation~(\ref{eq: Newton}), which is strongly nonlinear in essence as it can account accurately of multistreams, acts as a resummation of the LPT acceleration: it is expected to provide a more accurate prediction of the gravitational force field than the second time derivative of the LPT displacement. However, as long as the LPT series converges, we expect the higher-order LPT acceleration to converge to the force given by Eq.~(\ref{eq: Newton}) in the single stream regime, and this property will turn out to be useful even in the multistream regime when estimating the gravitational force orthogonal to the shell-crossing direction (coplanar with the pancake) by using Eq.~(\ref{eq: LPT acc multi}). 
\section{Analytical predictions for the gravitational force}
\label{sec: formula Fx}
In this section, we aim to compute the gravitational force shortly after the first shell-crossing in three-dimensional space. As detailed in Sect.~\ref{sec: setup problem}, we restrict to the formation of a symmetric pancake seeded by a locally axisymmetric motion. The calculation of the component of the force along the shell-crossing direction is the most challenging. However, after Taylor expanding the Lagrangian displacement field around the singularity just after shell-crossing, it turns out to be very similar to the pure one-dimensional case already treated in \citet{Gurevich_1995,2015MNRAS.446.2902C,2017MNRAS.470.4858T,2021MNRAS.505L..90R}. In particular it involves the resolution of a three-value problem related to the three flows inside the proto-pancake, as detailed in Sect.~\ref{sec: cubic eq}. The expression for the force along the shell-crossing direction is given in Sect.~\ref{sec: Fx formula}. In this subsection, we also argue that the force field in the transverse direction should not be significantly affected by the multi-stream nature of the flow, which will allow us to estimate it directly as the second time derivative of the displacement estimated with high-order LPT.

\subsection{Main assumptions}
\label{sec: setup problem}
\begin{figure*}
\centering
\includegraphics[width=\textwidth]{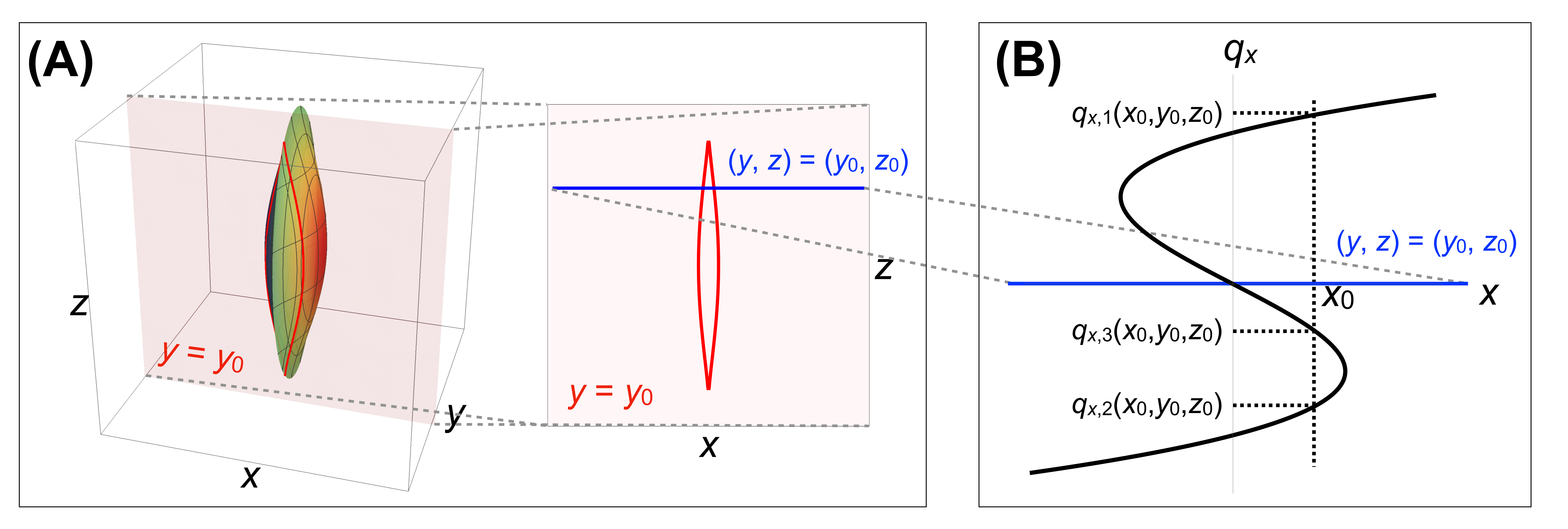}
\caption{
(A) {\em Left:} schematic representation of a three-dimensional Eulerian caustics shortly after shell-crossing along $x$-axis direction, together with a two-dimensional slice with $y=y_{0}$ (light-red plane). {\em Right:} the intersection of the caustic surface with the slice (red curve), also shown in the left panel.  (B) Schematic representation of the $x$-component of the Lagrangian coordinate $q_{x}$ as a function of the Eulerian coordinate $x$ for fixed $(y_{0},z_{0})$ (solid blue line). Given $x_{0}$, the solution of the three-value problem $x(q_x)=x_0$ is given by $q_{x,n}(x_{0},y_{0},z_{0})$ for $n=1$, $2$, and $3$, as in Eq.~(\ref{eq: Cardano}).
}
\label{fig: schematic}
\end{figure*}

In what follows, the calculations are all performed in 3D space, but the extension to 2D is straightforward by ignoring or setting to zero all the contributions depending on $z$. We also suppose that the first shell-crossing takes place at the origin, $\bm{q} = \bm{x} = \bm{0}$, along the $x$-axis direction, and also that the system exhibits locally axisymmetric dynamics. This setup, illustrated by Fig.~\ref{fig: schematic},  seemingly appears to be very particular, but locally represents the expected motion around high peaks of a Gaussian random field~\citep[see e.g.,][]{1986ApJ...304...15B}.

Axisymmetric dynamics translates as follows on the displacement $\bm{\Psi}(\bm{q})$:
\begin{align}
\Psi_{x}(q_x,q_y,q_z) &= \Psi_{x}(q_x,-q_y,q_z)  
= \Psi_{x}(q_x,q_y,-q_z) \notag \\
&
= - \Psi_{x}(-q_x,q_y,q_z) , \label{eq:sym1} \\
\Psi_{y}(q_x,q_y,q_z) &= \Psi_{y}(q_x,q_y,-q_z) 
= \Psi_{y}(-q_x,q_y,q_z) \notag \\
&
= - \Psi_{y}(q_x,-q_y,q_z) , \\
\Psi_{z}(q_x,q_y,q_z) &= \Psi_{z}(-q_x,q_y,q_z) 
= \Psi_{z}(q_x,-q_y,q_z) \notag \\
&
= - \Psi_{z}(q_x,q_y,-q_z) . \label{eq:sym3}
\end{align}
Here and hereafter, we omit the time dependence in the notations. Expanding these functions around the origin $\bm{q}=0$, we have
\begin{align}
\Psi_{x}(\bm{q}) &= \sum_{i,j,k=0} \frac{\psi_{2i+1\,2j\,2k}}{(2i+1)!(2j)!(2k)!} \, q^{2i+1}_{x}q^{2j}_{y}q^{2k}_{z} , \label{eq: Psi x expansion} \\
\Psi_{y}(\bm{q}) &= \sum_{i,j,k=0} \frac{\psi_{2i\,2j+1\,2k}}{{(2i)!(2j+1)!(2k)!}} \, q^{2i}_{x}q^{2j+1}_{y}q^{2k}_{z} , \\
\Psi_{z}(\bm{q}) &= \sum_{i,j,k=0} \frac{\psi_{2i\,2j\,2k+1}}{{(2i)!(2j)!(2k+1)!}} \, q^{2i}_{x}q^{2j}_{y}q^{2k+1}_{z} ,
 \label{eq: Psi z expansion}
\end{align}
with $\psi_{i\,j\,k}$ being some functions of time.  Substituting Eqs.~(\ref{eq: Psi x expansion})--(\ref{eq: Psi z expansion}) into Eq.~(\ref{eq: def Psi}), and neglecting $O(q^{4})$ and higher order terms, we obtain
\begin{align}
x(\bm{q}) &\simeq (1+\psi_{100})\, q_{x} + \frac{1}{2}\left( \psi_{120}\, q_{y}^{2} + \psi_{102}\, q_{z}^{2}\right) q_{x} + \frac{1}{6} \psi_{300}\, q_{x}^{3} , \label{eq: xq 1}\\
y(\bm{q}) &\simeq (1+\psi_{010})\, q_{y} + \frac{1}{2}\left( \psi_{012}\, q_{z}^{2} + \psi_{210}\, q_{x}^{2}\right) q_{y} + \frac{1}{6} \psi_{030}\, q_{y}^{3} , \label{eq: yq 1}\\
z(\bm{q}) &\simeq (1+\psi_{001})\, q_{z} + \frac{1}{2}\left( \psi_{201}\, q_{x}^{2} + \psi_{021}\, q_{y}^{2}\right) q_{z} + \frac{1}{6} \psi_{003}\, q_{z}^{3} . \label{eq: zq 1}
\end{align}
While this local representation of the motion is minimal, it remains accurate shortly after collapse as illustrated by Appendix~\ref{app: x_qx} . 

We now write the necessary conditions that the coefficients in Eqs.~(\ref{eq: xq 1})--(\ref{eq: zq 1}) must satisfy for a pancake to exist near the origin. It is important to note that these conditions do not necessarily imply that a halo subsequently forms, this would require more restrictive constraints. 

Since we consider a system in which first shell-crossing just took place along the $x$-direction at $\bm{q}=\bm{0}$, we impose
\begin{align}
\frac{\partial x(\bm{0})}{\partial q_{x}} &\equiv -h = 1+\psi_{100} < 0 , \quad 0 < h \ll 1, \label{eq: condition qx}\\
\frac{\partial y(\bm{0})}{\partial q_{y}} &= 1+\psi_{010} > 0 , \label{eq: condition qy} \\
\frac{\partial z(\bm{0})}{\partial q_{z}} &= 1+\psi_{001} > 0 . \label{eq: condition qz}
\end{align}
Additional constraints can be obtained from the expression of the Jacobian determinant, $J = \det{\partial\bm{x}/\partial\bm{q}}$ at leading order in $\bm{q}$, which reads 
\begin{align}
J \simeq \frac{1}{2}(1+\psi_{010})(1+\psi_{001})\left( -2h + \psi_{120}q^{2}_{y} + \psi_{102}q^{2}_{z} + \psi_{300}q_{x}^{2}\right). \label{eq: leading J}
\end{align}
From catastrophe theory \citep[see, e.g.,][and references therein]{2014MNRAS.437.3442H,2018JCAP...05..027F}, the caustic surface should be an ellipsoid outside of which the Jacobian determinant must have a positive value $J>0$, so we impose
\begin{align}
\psi_{120} >0 ,~~~ \psi_{102} >0 ,~~~ \psi_{300} >0 . 
\end{align}

Finally, the smallness of the $h$ parameter induces an additional simplification of Eqs.~(\ref{eq: xq 1})--(\ref{eq: zq 1}) if one supposes that Lagrangian coordinates are restricted to lie in the neighbourhood of the pancake, i.e. $q_{x} \sim q_{y} \sim q_{z} \sim \mathcal{O}(h^{1/2})$ from Eq.~(\ref{eq: leading J}). In this case, one realizes that, when examining Eqs.~(\ref{eq: xq 1})--(\ref{eq: zq 1}), 
\begin{align}
x(\bm{q}) &= \mathcal{O}(h^{3/2}) , \label{eq: xq 2}\\
y(\bm{q}) &\simeq (1+\psi_{010}) q_{y} + \mathcal{O}(h^{3/2}) , \label{eq: yq 2}\\
z(\bm{q}) &\simeq (1+\psi_{001}) q_{z} + \mathcal{O}(h^{3/2}) , \label{eq: zq 2}
\end{align}
which implies $|x|\sim \mathcal{O}(h^{3/2}) \ll |y|,\, |z| \sim \mathcal{O}(h^{1/2})$, a signature of the pancake nature of the system: the extension of the caustic region along the $x$-axis is asymptotically infinitely smaller than that along the other axes in the limit $h\to 0$, as illustrated by panel (A) of Fig.~\ref{fig: schematic}. Accordingly, inside and in the vicinity of the caustic region, we can ignore the higher-order terms in the expressions of $y$ and $z$, and reduce Eqs.~(\ref{eq: xq 1})--(\ref{eq: zq 1}) to
\begin{align}
x(\bm{q}) &\simeq (1+\psi_{100}) q_{x} + \frac{1}{2}\left( \psi_{120}\, q_{y}^{2} + \psi_{102}\, q_{z}^{2}\right) q_{x} + \frac{1}{6} \psi_{300}\, q_{x}^{3} , \label{eq: xq 3}\\
y(\bm{q}) &\simeq (1+\psi_{010}) q_{y} , \label{eq: yq 3}\\
z(\bm{q}) &\simeq (1+\psi_{001}) q_{z} . \label{eq: zq 3}
\end{align}
Eqs.~(\ref{eq: xq 3})--(\ref{eq: zq 3}) represent our starting point to derive the $x$-component of the gravitational force inside a pancake. 

\subsection{The three value problem}
\label{sec: cubic eq}
Despite its apparent simplicity, Eq.~(\ref{eq: Newton}) is not easily exploitable when it comes to estimate the gravitational force analytically, even with as simple expressions as Eqs.~(\ref{eq: xq 3})--(\ref{eq: zq 3}). Indeed, although the multi-stream nature of the flow does not appear explicitly in the integral (\ref{eq: Newton}), we shall see in Sect.~\ref{sec: Fx formula} that the calculation of the $x$-component of the gravitational force still requires solving the three-valued problem implicit in  Eqs.~(\ref{eq: xq 3})--(\ref{eq: zq 3}), that is finding $\bm{q}$ given $\bm{x}$. 

From Eqs.~(\ref{eq: yq 3}) and (\ref{eq: zq 3}), we trivially obtain
\begin{align}%
q_{y} = \frac{y}{1+\psi_{010}} ,~~~ q_{z} = \frac{z}{1+\psi_{001}} . \label{eq:qyqx}
\end{align}
The calculation of $q_{x}(\bm{x})$ is more complex because it requires solving the following cubic equation, as illustrated by panel (B) of Fig.~\ref{fig: schematic},
\begin{align}%
q_{x}^{3} + 3A(y,z)\, q_{x} +2B(x) = 0 , \label{eq: cubic}
\end{align}
where we defined
\begin{align}%
& A(y,z) = \frac{1}{\psi_{300}}\left( -2h +  \psi_{120}\left( \frac{y}{1+\psi_{010}}\right)^{2} + \psi_{102}\left( \frac{z}{1+\psi_{001}} \right)^{2} \right) , \label{eq:Aconst}\\
& B(x) = - \frac{3x}{\psi_{300}} .\label{eq:Bconst}
\end{align}
The roots of cubic Eq.~(\ref{eq: cubic}) are given by \citep[see e.g.,][]{1972hmfw.book.....A}
\begin{align}%
q_{x,n}(\bm{x}) &=
\omega^{n-1}\frac{-A(y,z)}{\left( \sqrt{D(\bm{x})} - B(x) \right)^{1/3}}+ \omega^{4-n} \left( \sqrt{D(\bm{x})} -B(x) \right)^{1/3}
,
\label{eq: Cardano}
\end{align}
for $n=1$, $2$, and $3$. Here, the factor $\omega$ is one of the complex cubic roots of unity, i.e., $\omega = (-1\pm i\sqrt{3})/2$, which are the solutions of $\omega^{2}+\omega+1=0$. We define the discriminant $D(\bm{x})$, which determines the properties of the roots (\ref{eq: Cardano}), by \citep[e.g.,][]{1972hmfw.book.....A}
\begin{align}%
D(\bm{x}) &= (A(y,z))^{3} + (B(x))^{2} .
\label{eq: discriminant}
\end{align}
First note that the equation $D(\bm{x}) = 0$ defines the caustics surfaces in Eulerian space: 
\begin{align}%
\Biggl[ -2h
+ \psi_{120}\left( \frac{y}{1+\psi_{010}} \right)^{2} 
+ \psi_{102}\left( \frac{z}{1+\psi_{001}} \right)^{2}
\Biggr]^{3}
+ 9\psi_{300}x^{2} = 0. \label{eq:causticsurface}
\end{align}
According to the sign of $D(\bm{x})$, the solutions of the cubic equation (\ref{eq: cubic}) can then be classified as follows. If $D(\bm{x})<0$, we are inside the multi-stream region delimited by the caustic surfaces: the cubic equation has three real solutions, with $q_{x,2}<q_{x,3}<q_{x,1}$ and $q_{x,2} < 0 < q_{x,1}$. If $D(\bm{x})>0$, we are outside the multi-stream region: only either $q_{x,1}$ or $q_{x,2}$ is real, and the other solutions are complex conjugate to each other.

As a final note, from Eq.~(\ref{eq:causticsurface}), the maximum values of the Eulerian coordinates of caustics along the $x$-, $y$-, and $z$-axes are given by
\begin{align}%
x_{\rm max} &= \sqrt{\frac{8h^{3}}{9\psi_{300}}} , \label{eq: xmax}\\
y_{\rm max} &= (1+\psi_{010})\sqrt{\frac{2h}{\psi_{120}}} , \label{eq: ymax}\\
z_{\rm max} &= (1+\psi_{001})\sqrt{\frac{2h}{\psi_{102}}} , \label{eq: zmax}
\end{align}
Once again, $x_{\rm max} \ll y_{\rm max}, z_{\rm max}$ in the limit $h \to 0$.
\subsection{The Force field inside a pancake}
\label{sec: Fx formula}
We now present the main results of this paper. Focusing on the dynamics just after the first shell crossing along $x$-direction, we first derive an analytical formula for the $x$-component of the gravitational force after collapse. Next, we discuss orthogonal components and argue that they can be directly approximated by the high-order LPT acceleration.

By zooming in very closely on the multi-stream region, we notice that the structure of the density field becomes almost one dimensional when $h \rightarrow 0$. This follows from the fact the size of the pancake along the $x$-axis is much smaller than along orthogonal directions (see Eqs.~\ref{eq: xq 2}--\ref{eq: zq 2} and also panel A of Fig.~\ref{fig: schematic}). Consequently, given the Eulerian position $(x,y_{0},z_{0})$ for $y_{0}$ and $z_{0}$ fixed, solving the Poisson equation (\ref{eq: delta phi}) is asymptotically reduced to a one-dimensional problem along $x$ axis, as schematically shown in panel (B) of Fig.~\ref{fig: schematic}. In this case, we can neglect local variations of the density along the $y$ and $z$-directions, and Eq.~(\ref{eq: delta phi}) reduces to the following one-dimensional Poisson equation for the $x$-coordinate of the gravitational acceleration, 
\begin{align}
\frac{{\rm d}^{2}\phi_{\rm p}(x,y_{0},z_{0})}{{\rm d}x^{2}} &= 4\pi G\bar{\rho}a^{2} \, (1+\delta(x,y_{0},z_{0})) , \notag \\
&= 4\pi G\bar{\rho}a^{2} \, \frac{1}{(1+\psi_{010})(1+\psi_{001})}\left| \frac{\partial x}{\partial q_{x}} \right|^{-1} ,
 \label{eq: delta phi 1D}
\end{align}
where, in the second equality, we have used Eq.~(\ref{eq: delta-J}) with Eqs.~(\ref{eq: xq 3})--(\ref{eq: zq 3}). 

To solve this equation, we follow the footsteps of~\citet{2015MNRAS.446.2902C,2017MNRAS.470.4858T,2021MNRAS.505L..90R} and employ a Green function approach to  derive the $x$-component of the force in the multi-stream region, $F_{x}(\bm{x}_{0}) \simeq  -{\rm d}\phi_{\rm p}(\bm{x}_{0})/{\rm d}x$ with $\bm{x}_{0} = (x_{0},\, y_{0},\, z_{0})$ (inside the pancake, $x \sim h^{3/2}$, and the background contribution ${\bar \phi}$ is negligible):
\begin{widetext}
\begin{align}
F_{x}(\bm{x}_{0})
&\simeq \int{\rm d}x\, \frac{4\pi G\bar{\rho}a^{2}}{(1+\psi_{010})(1+\psi_{001})}
\sum_{n=0,1,2} \left| \frac{\partial x}{\partial q_{x,n}} \right|^{-1}
\frac{1}{2}\Bigl[ \Theta_{\rm H}(x-x_{0}) - \Theta_{\rm H}(x_{0}-x) \Bigr] , \notag \\
&= \int{\rm d}q_{x}\, \frac{4\pi G\bar{\rho}a^{2}}{(1+\psi_{010})(1+\psi_{001})}
\frac{1}{2}\left[ \Theta_{\rm H}\left( x\left(q_{x},\frac{y_{0}}{1+\psi_{010}},\frac{z_{0}}{1+\psi_{001}}\right)-x_{0} \right) - \Theta_{\rm H}\left( x_{0}-x\left(q_{x},\frac{y_{0}}{1+\psi_{010}},\frac{z_{0}}{1+\psi_{001}}\right) \right) \right] , \notag \\
&= - \frac{4\pi G\bar{\rho}a^{2}}{(1+\psi_{010})(1+\psi_{001})} \Bigl[ q_{x,1}(\bm{x}_{0}) + q_{x,2}(\bm{x}_{0}) - q_{x,3}(\bm{x}_{0})\Bigr]
\qquad \mbox{(in the multi-stream region)}
 , \label{eq: our formula multi}
\end{align}
\end{widetext}
where $\Theta_{\rm H}$ represents the Heaviside step function, and the quantities $q_{x,n}$ are the three-value problem solutions given in Eq.~(\ref{eq: Cardano}). In the first line of Eq.~(\ref{eq: our formula multi}), we have dropped the negligible contributions in the one-dimensional Green function that are not proportional to the Heaviside step function. 

Interestingly, one can also write, asymptotically,
\begin{align}
  F_{x}(\bm{x}_{0}) \simeq \Gamma_x[q_{x,1}(\bm{x}_{0})] + \Gamma_x[q_{x,2}(\bm{x}_{0})] - \Gamma_x[q_{x,3}(\bm{x}_{0})], \label{eq:acc1D}
\end{align}
where $\Gamma_x$, defined in Eq.~(\ref{eq: acc each flow}), is the $x$-coordinate of the LPT acceleration (computed by ignoring shell-crossing). This approximation will be tested in Sect.~\ref{sec: comparison}. 

Eq.~(\ref{eq: our formula multi}) applies only to the multi-stream region. However, as mentioned above, in the single-stream region, either $q_{x,1}$ or $q_{x,2}$ is real while the two other values are complex conjugate to each other. Taking the real part of right hand side of Eq.~(\ref{eq: our formula multi}), the unphysical contribution arising from the complex conjugate solutions vanishes, leaving only the physical contribution from one real solution\footnote{Equivalently, we can rewrite Eq.~(\ref{eq: our formula multi}) as follows:
\begin{align}
F_{x}(\bm{x}_{0}) = - \frac{4\pi G\bar{\rho}a^{2}}{(1+\psi_{010})(1+\psi_{001})} \Bigl[ q_{x,1}(\bm{x}_{0}) + q_{x,2}(\bm{x}_{0}) - \left(q_{x,3}(\bm{x}_{0})\right)^{*}\Bigr]
\end{align}
where the asterisk denotes the complex conjugate.
}:
\begin{align}
F_{x}(\bm{x}_{0})
&\simeq - \frac{4\pi G\bar{\rho}a^{2}}{(1+\psi_{010})(1+\psi_{001})} {\rm Re}{\Bigl[ q_{x,1}(\bm{x}_{0}) + q_{x,2}(\bm{x}_{0}) - q_{x,3}(\bm{x}_{0})\Bigr]}
 . \label{eq: our formula}
\end{align}
Here, ``${\rm Re}$'' denotes the real part.

Eq.~(\ref{eq: our formula}) is the most important formula in our work. Together with the solution of the cubic equation~(\ref{eq: Cardano}), it consists of a simple algebraic form for the $x$-component of the force in the vicinity of the pancake, inside {\em and} outside it.  This formula is the natural extension to three dimensions of the one dimensional calculations of~\citet{Gurevich_1995,2015MNRAS.446.2902C,2017MNRAS.470.4858T,2021MNRAS.505L..90R}. 

The components $F_y$ and $F_z$ of the force orthogonal to the shell-crossing direction, hence coplanar with the pancake, should, on the other hand, remain quite insensitive to the effects of shell-crossing, as we will show with the numerical tests performed in next sections. This means that if the LPT displacement field is entirely sourced by gravity, its second time derivative should provide a very good approximation of the local gravitational acceleration, even slightly beyond shell crossing. As a consequence, the acceleration directly derived from the high-order LPT solution (with no Taylor expansion at third order in $\bm{q}$ space) should provide a good approximation for $F_y$ and $F_z$. Of course this is true only if one remains within the convergence radius of the LPT series, which is finite in the general case~\citep[e.g.,][]{2014JFM...749..404Z,2015MNRAS.452.1421R,2021MNRAS.501L..71R}. To overcome the three-value problem, the acceleration can be obtained with the weighted average (\ref{eq: LPT acc multi}) over the different flows inside the pancake, even though this is not absolutely necessary, since it is expected, in the vicinity of the pancake, that $F_y[\bm{x}(\bm{q}_1)]\simeq F_y[\bm{x}(\bm{q}_2)] \simeq F_y[\bm{x}(\bm{q}_3)]$ and similarly for the $z$-component of the force: this is due to the fact that the vectors $\bm{q}_1$, $\bm{q}_2$ and $\bm{q}_3$ are nearly equal. 
\section{Examination of the accuracy of the formulas}
\label{sec: analytical ex}
In this section, we test the validity and the accuracy of our prescription for calculating the post-collapse force. To this end, we assume Einstein-de Sitter cosmology and consider simplified initial conditions composed of two or three crossed sine waves, following the early investigations of~\citet{1991ApJ...382..377M,1995ApJ...441...10M} and our previous works~\citep[][]{2018PhRvL.121x1302S}. We present these initial conditions in Sect.~\ref{sec:sine_ini}, where we also give practical details on the way the analyses are performed. Sect.~\ref{sec: analytical study Fx} focuses on the force along the shell-crossing direction. We test, at different orders of the perturbative development, the accuracy of Eq.~(\ref{eq: our formula}) against the exact solution obtained by direct resolution of Poisson equation. This is followed in Sect.~\ref{sec: analytical study Fy} by a comparison, in the transverse direction, of the LPT acceleration to the exact force. All these analyses are performed very shortly after collapse to make sure that assumptions of Sect.~\ref{sec: setup problem} are verified. One indeed expects increasing discrepancies with time between the exact solution and the approximations of the dynamics, as examined in Sect.~\ref{sec: time}. 

Throughout this section, we use the LPT solution as a proxy for the ``exact'' Eulerian position.  To calculate the ``exact'' force field, we simply inject the LPT solution into Eq.~(\ref{eq: Newton}) and numerically perform the two- or three-dimensional integrals.\footnote{In numerically integrating Eq.~(\ref{eq: Newton}), we use the {\tt NIntegrate} function in {\tt Mathematica} with the options {\tt \{MaxRecursion -> 100 (10000), PrecisionGoal -> 5 (8), Method -> \{Automatic, "SymbolicProcessing" -> 0\}\}} for $F_{x}$ ($F_{y,z}$). In this instance, the computation of the results depicted in Figs.~\ref{fig: Fx merge} or \ref{fig: Fy merge}, on an 8-core CPU laptop, takes several hours.}
The comparisons to simulations performed in Sect.~\ref{sec: comparison} will show that this proxy turns to provide a quite accurate description of the true displacements at the times we consider. 
\subsection{Two and three sine waves initial conditions: practical implementation}
\label{sec:sine_ini}
\begin{table*}
\centering
\begin{tabular}{ccccccccc}
\hline
Designation & $\epsilon_{\rm 2D}$ or $\bm{\epsilon}_{\rm 3D}$ & $\epsilon_x$
& $n_{\rm g}$ & $n_{\rm s}$
& $a^{\rm LPT}_{\rm sc}$ & $a_{\rm sc}$ & $a_{\rm sim}$
& $\Delta_{\rm sim} \equiv (a_{\rm sim}-a_{\rm sc})/a_{\rm sc}$ \\
\hline
{\it Quasi 1D}\\
Q1D-2SIN & 1/6 & -18 & 2048 & 2048 & 0.05279 & 0.05285 & 0.05402 & 0.02219\\
Q1D-3SIN & (1/6, 1/8) & -24 & 512 & 256 & 0.03814 & 0.03832 & 0.03907 & 0.01974\\
\hline
{\it Anisotropic}\\
ANI-2SIN & 2/3 & -18 & 2048 & 2048 & 0.04534 & 0.04545 & 0.04601 & 0.01230\\
ANI-3SIN & (3/4, 1/2) & -24 & 512 & 512 & 0.02912 & 0.02919 & 0.03003 & 0.02890\\
\hline
{\it Axial-symmetric}\\
SYM-2SIN & 1 & -18 & 2048 & 2048 & 0.04087 & 0.04090 & 0.04101 & 0.002717\\
SYM-3SIN & (1, 1) & -18 & 512 & 512 & 0.03236 & 0.03155 & 0.03201 & 0.01446\\
\hline
\end{tabular}
\caption{Parameters of the runs performed with \texttt{ColDICE}~\citep{2016JCoPh.321..644S}. The first column indicates the designation of the run. The second column corresponds to the relative amplitudes of the initial sine waves, namely, $\epsilon_{\rm 2D} = \epsilon_{y}/\epsilon_{x}$ and $\bm{\epsilon}_{\rm 3D} = (\epsilon_{y}/\epsilon_{x}, \epsilon_{z}/\epsilon_{x})$ for two and three sine waves, respectively. The third column gives the value of $\epsilon_x$.
The fourth and fifth columns indicate the resolutions of the numerical simulations: the spatial resolution of the grid used to solve the Poisson equation, and the spatial resolution of the mesh of vertices used to construct the initial tessellation, respectively. In the sixth column, $a^{\rm LPT}_{\rm sc}$ is the shell-crossing time estimated by 40LPT for the two-sine waves initial conditions and by 15LPT for the three-sine waves initial conditions. The seventh, eighth, and ninth columns indicate the scale factor $a_{\rm sc}$ at shell-crossing measured in the simulations (see Appendix~A1 in STC), the scale factor of the output time used for the analyses performed shortly after collapse in the simulations, and the fractional difference between $a_{\rm sc}$ and $a_{\rm sim}$, respectively.
}
\label{tab: initial conditions}
\end{table*}
We consider initial conditions seeded by two or three crossed sine waves in a periodic box $[-L/2, L/2[$ in which the initial displacement field at initial time $t_{\rm ini}$ is expressed by~(see STC)
\begin{align}
\Psi_{i}^{\rm ini}(\bm{q}, t_{\rm ini}) =\frac{L}{2\pi}\,D_+(t_{\rm ini})\,\epsilon_{i} \sin\left( \frac{2\pi}{L}q_{i}\right) ,
\label{eq:init_psi}
\end{align}
with $\epsilon_{i} < 0$ and $|\epsilon_x| \geq |\epsilon_y| \geq |\epsilon_z|$. The initial density field, $\delta^{\rm ini} \simeq -\bm{\nabla}_{q}\cdot\bm{\Psi}^{\rm ini} = D_+(t_{\rm ini}) \sum_{i}|\epsilon_{i}| \cos\left( 2\pi/L\,q_{i}\right)$, presents a small peak at the origin. Subsequently, mass elements fall towards the overdense central region, and shell-crossing takes place at the origin. The initial time, $t_{\rm ini}$, is set to satisfy $D_+(t_{\rm ini})|\epsilon_{i}|\le 0.012 \ll 1$, so that the fastest growing mode approximation is accurate~\citep{2018PhRvL.121x1302S}. We note that in the Einstein-de Sitter universe, the growth factor is simply proportional to the scale factor and we have $f=1$. Hence, we shall hereafter use the scale factor to describe the time rather than $D_+$.

In this setup, the dynamics is determined by the ratios $\epsilon_{\rm 2D} = \epsilon_{y}/\epsilon_{x}$ and $\bm{\epsilon}_{\rm 3D}=(\epsilon_{y}/\epsilon_{x}, \epsilon_{z}/\epsilon_{x})$, respectively for two and three sine waves initial conditions (STC). We consider three qualitatively different initial conditions as detailed in Table~\ref{tab: initial conditions}: quasi one-dimensional with $|\epsilon_x| \gg |\epsilon_{y,z}|$ (Q1D), anisotropic with $|\epsilon_x| > |\epsilon_y| > |\epsilon_z|$ (ANI), and axial-symmetric with $|\epsilon_x| = |\epsilon_y| = |\epsilon_z|$ (SYM).  The Q1D and ANI initial conditions are the primary targets of our analyses because first shell crossing occurs only along the $x$-direction, satisfying the assumptions of Sect.~\ref{sec: setup problem}. On the other hand, for the SYM case, since shell crossing takes place simultaneously along two or three axial directions, the prescriptions proposed in Sect.~\ref{sec: Fx formula} to approximate the force field becomes improper. However, the SYM case might provide insight into rare primordial haloes and remains mathematically interesting. In this case, we will only explore numerically in Sect.~\ref{sec: comparison} the force field by comparing predictions of higher-order LPT combined with the Green function approach (Eq.~\ref{eq: Newton}) to measurements in the simulations. 

Sine waves initial conditions (\ref{eq:init_psi}) correspond to a low order trigonometric polynomial, which makes LPT calculations relatively cheap and  allows us to reach $40^{\rm th}$ and $15^{\rm th}$ order for the two- and three-dimensional case, respectively  (see Sect.~2 in STC for the explicit procedure).\footnote{The high-order LPT solutions can be provided upon request as a {\tt Mathematica} notebook.} 

Having obtained the higher-order LPT solutions, we analytically predict the force along the shell-crossing direction using Eq.~(\ref{eq: our formula}), by expanding the LPT solutions around the origin up to third order in terms of Lagrangian coordinates to have the expressions for the coefficients $\psi_{ijk}$ in Eqs.~(\ref{eq: Psi x expansion})--(\ref{eq: Psi z expansion}) hence in formula~(\ref{eq: our formula}). Also, we analytically predict the force orthogonal to the shell-crossing direction by directly using the LPT acceleration (without Taylor expanding it in $\bm{q}$) and averaging it (numerically) over different flows at same Eulerian position $\bm{x}$ using Eq.~(\ref{eq: LPT acc multi}). 

Because the collapse time, traced here by the value of the expansion factor $a^{(n)}_{\rm sc}$, decreases when perturbative order $n$ augments (see, e.g., STC), LPT predictions are in practice synchronized to their own respective shell-crossing times. Then the system is evolved beyond collapse by a very small amount of time, i.e.,
\begin{align}
  a^{(n)} = a^{(n)}_{\rm sc} \left( 1+ \Delta \right),\label{eq:deltadef}
\end{align}
where $\Delta$ is a small parameter, such that the assumptions of Sect.~\ref{sec: setup problem} are valid, in particular $h \simeq \Delta\, |\partial v_{x}/\partial q_{x}/(aH)|_{\bm{q}=\bm{0},\, t=t_{\rm sc}} \ll 1$. In practice, we shall take $\Delta = 0.001$ in Secs.~\ref{sec: analytical study Fx} and \ref{sec: analytical study Fy}, while it will be allowed to be larger in Sect.~\ref{sec: time}. 

As a final note, we truncate the integral in Eq.~(\ref{eq: Newton}) to a finite interval $[-q_{\rm max}, q_{\rm max}]$ in each dimension, setting $q_{\rm max} = L/2$ for $F_{x}$ and $q_{\rm max} = 10L$ for $F_{y,z}$ to insure convergence of the integral, as studied in Appendix~\ref{app: different G}. 

In the rest of the paper, we will use the following units: a box size $L = 1$ and an inverse of the Hubble parameter at the present time $H_{0} =1$ for the dimensions of length and time, respectively. We will also present the normalised force $\tilde{\bm{F}}(\bm{x})$ given by
\begin{align}%
\tilde{\bm{F}}(\bm{x}) = \frac{\bm{F}(\bm{x})}{4\pi G a^{2}\bar{\rho}} .
\end{align}

\subsection{Force along the shell-crossing direction, $F_{x}$}
\label{sec: analytical study Fx}
\begin{figure}
\centering
\includegraphics[width=\columnwidth]{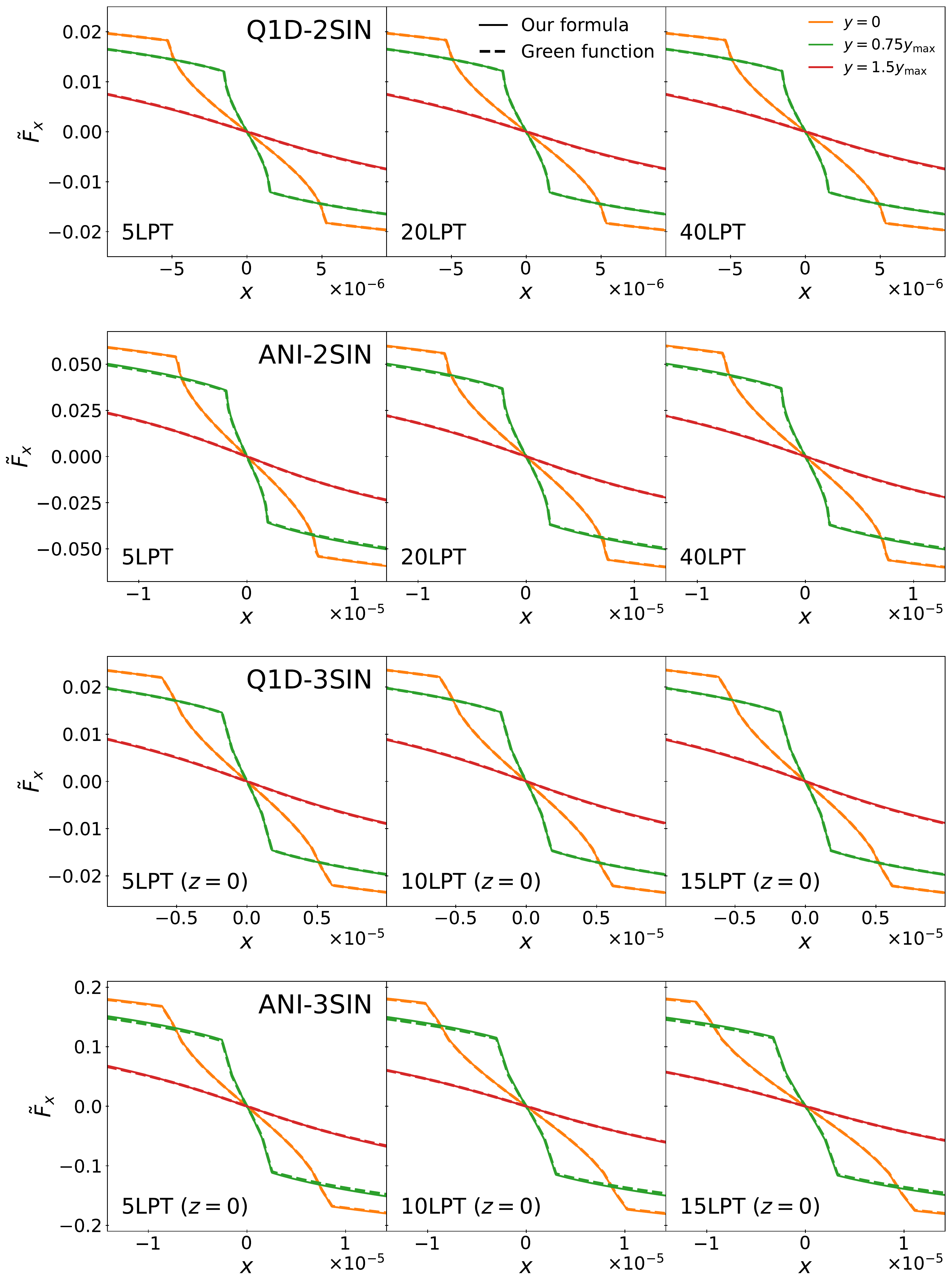}
\caption{
$x$-component of force as a function of $x$ inside and in the vicinity of a pancake seeded by three sine waves in the context of LPT dynamics, with  $-1.8\leq x/x_{\rm max} \leq 1.8$, where $x_{\rm max}$ is the maximum extension of the caustic region along $x$ axis as given by Eq.~(\ref{eq: xmax}) in our approximate formalism. The output time is set to be $\Delta = 0.001$ with synchronisation (see Eq.~(\ref{eq:deltadef})).
On each panel, the curves of various colours correspond to different values of $y$ (and $z=0$ in the 3D case), namely $y/y_{\rm max} = 0$ (orange), $0.75$ (green), and $1.5$ (red, outside the multistream region), where $y_{\rm max}$ is the maximum extension of the caustic region along $y$ axis (Eq.~\ref{eq: ymax}).  From top to bottom, we consider Q1D-2SIN, ANI-2SIN, Q1D-3SIN, and ANI-3SIN. From left to right, the predictions are made with 5, 20 (10), and 40LPT (15LPT) for two-sine (three-sine) waves initial conditions, respectively. The dashed and solid lines represent, respectively, the ``exact'' force given by Eq.~(\ref{eq: Newton}) and the analytic prediction~(\ref{eq: our formula}).
}
\label{fig: Fx merge}
\end{figure}
Fig.~\ref{fig: Fx merge} displays, just after shell-crossing, the $x$-component of the force $F_{x}$ as a function of $x$ in the vicinity of the pancake for various values of $y$. Perturbative order increases from left to right, while initial conditions are given, from top to bottom, by Q1D-2SIN, ANI-2SIN, Q1D-3SIN, and ANI-3SIN. 

We first notice the perfect agreement between the analytical formula (\ref{eq: our formula}) (solid curves) and the exact solution (\ref{eq: Newton}) (dashed curves), which fully justifies the mathematical relevance of the formalism developed in Sect.~\ref{sec: formula Fx}.  In particular, the sharp transition between the single-stream and the multi-stream region is perfectly described by Eq.~(\ref{eq: our formula}), with the variations with $y$ (and $z$, not shown here) accounted for correctly. The discontinuity observed on the derivative of the $x$-component of the force field is a typical signature of the presence of caustics, as found previously in the 1D case~\citep[see, e.g.,][]{Gurevich_1995,2015MNRAS.446.2902C}, as well as in caustic rings~\citep[see, e.g., Fig.~8 in][]{1999PhRvD..60f3501S}. 

Interestingly, the results do not strongly depend on the perturbation order in Fig.~\ref{fig: Fx merge}. This is because both the analytical formula (\ref{eq: our formula}) and the numerical calculation (\ref{eq: Newton}) rely on LPT displacement, which shows fast convergence behaviour with LPT order after synchronisation of shell-crossing times~(see Fig.~10 in STC).

Obviously, our approach works because parameter $\Delta$ in Eq.~(\ref{eq:deltadef}) is very small, which makes the pancake quasi one dimensional, independently of the values of $\epsilon_{2D}$ and $\bm{\epsilon}_{3D}$ as long as $\epsilon_y, \epsilon_z < \epsilon_x$. We will see in Sect.~\ref{sec: time} how the accuracy of the description deteriorates with increasing $\Delta$, that is with increasing time interval after collapse.
\subsection{Force orthogonal to the shell-crossing direction}
\label{sec: analytical study Fy}
\begin{figure}
\centering
\includegraphics[width=\columnwidth]{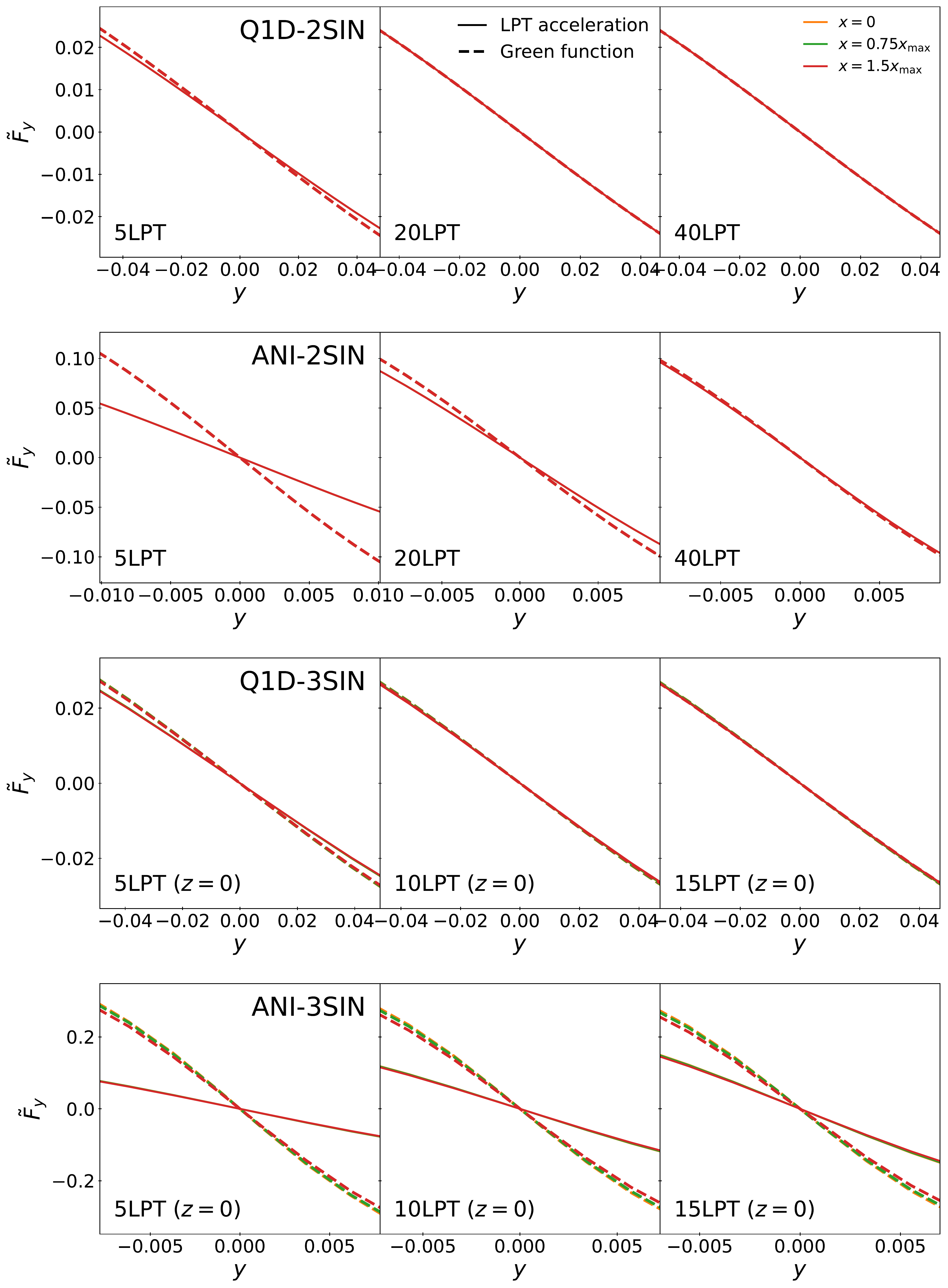}
\caption{
Same as Fig.~\ref{fig: Fx merge} but for the $y$-component of the force, $F_{y}$, in the range $-1.8\leq y/y_{\rm max} \leq 1.8$ for $x/x_{\rm max}=0$ (orange), $0.75$ (green), and $1.5$ (red).  The dashed and solid lines represent, respectively, the force given in Eq.~(\ref{eq: Newton}) and the LPT acceleration given in Eq.~(\ref{eq: LPT acc multi}). Due to the weak dependence of $F_{y}$ on the presence of the caustic and the very small range of values of $x$ considered, the curves for $x/x_{\rm max}=0$, $0.75$, and $1.5$ nearly perfectly overlap.
}
\label{fig: Fy merge}
\end{figure}
We now turn to the component $F_y$ of the force orthogonal to the shell-crossing direction (hence coplanar with the pancake) inside and in the vicinity of the pancake, as shown in Fig.~\ref{fig: Fy merge} very shortly after collapse (note that the plot of the $z$-component, $F_z$, in the 3D case, would give results very similar to $F_y$, so is not shown here). As already mentioned in Sect.~\ref{sec: Fx formula}, contrary to $F_x$, the  exact gravitational force field given by Eq.~(\ref{eq: Newton}), displayed as dashed curves, is not significantly affected by the presence of caustics,  with a weak dependence on $x$, $F_{y}\propto y$ and $F_{z}\propto z$ close to the origin, while $F_y$ and $F_z$ are locally even with respect to $z$ and $y$, respectively (not shown on the figures). In particular we found by comparing $F_{y,z}$ just before and after collapse ($\Delta=\pm 0.001$) that it hardly changes during this period of time. These results suggest that shell-crossing along the $x$-direction does not strongly affect the dynamics along other axes. This property allows us to still use the high-order LPT solution, that is the acceleration computed as the second time derivative of the LPT displacement, to describe the $y$ and $z$ components of the force as long as the LPT series converges, as shown by the solid curves in Fig.~\ref{fig: Fy merge} after averaging other the streams according to Eq.~(\ref{eq: LPT acc multi}). The solid curves in Fig.~\ref{fig: Fy merge} converge to the exact solution when increasing perturbative order $n$, except that $15^{\rm th}$ order is still insufficient for ANI-3SIN. Convergence is eased when approaching quasi one dimensional initial conditions (Q1D-2SIN and Q1D-3SIN), thanks to the much faster convergence of the LPT series in this case.  We also checked in Eq.~(\ref{eq: LPT acc multi}) that the LPT accelerations computed inside each stream are nearly the same. 

Another important finding is that the dashed curves are only weakly dependent on LPT order, which is a consequence of synchronisation of LPT predictions with their own collapse time. We shall see in Sect.~\ref{sec: comparison} that they actually agree very well with the measurements in {\tt ColDICE} simulations. As shown by STC, the density field, which stems from synchronized LPT displacements and sources the force field, is very well described by LPT predictions even of relatively low order. On the other hand, the solid curves in Fig.~\ref{fig: Fy merge} show that the LPT acceleration converges much more slowly than the LPT displacement, even after synchronisation. This is a rather obvious consequence of the fact that time derivatives give more weight than cumulative quantities to deviations from the exact solution, since these last ones increase with time. Similarly, STC found that convergence of LPT for velocities fields was slower than for density fields. Note thus that Eq.~(\ref{eq: Newton}) provides, at fixed LPT order, a much more accurate way to compute the gravitational acceleration than the second time derivative of the LPT displacement, as already mentioned in last paragraph of Sect.~\ref{sec: force}. In other words, it provides a resummation of the gravitational force that could be used to improve on LPT predictions even prior to shell crossing (i.e., even in the absence of synchronisation, see Fig.~\ref{fig: sim wo sync} in Sect.~\ref{sec: comparison}). 

\subsection{Time dependence}
\label{sec: time}

\begin{figure*}
\centering
\includegraphics[width=\textwidth]{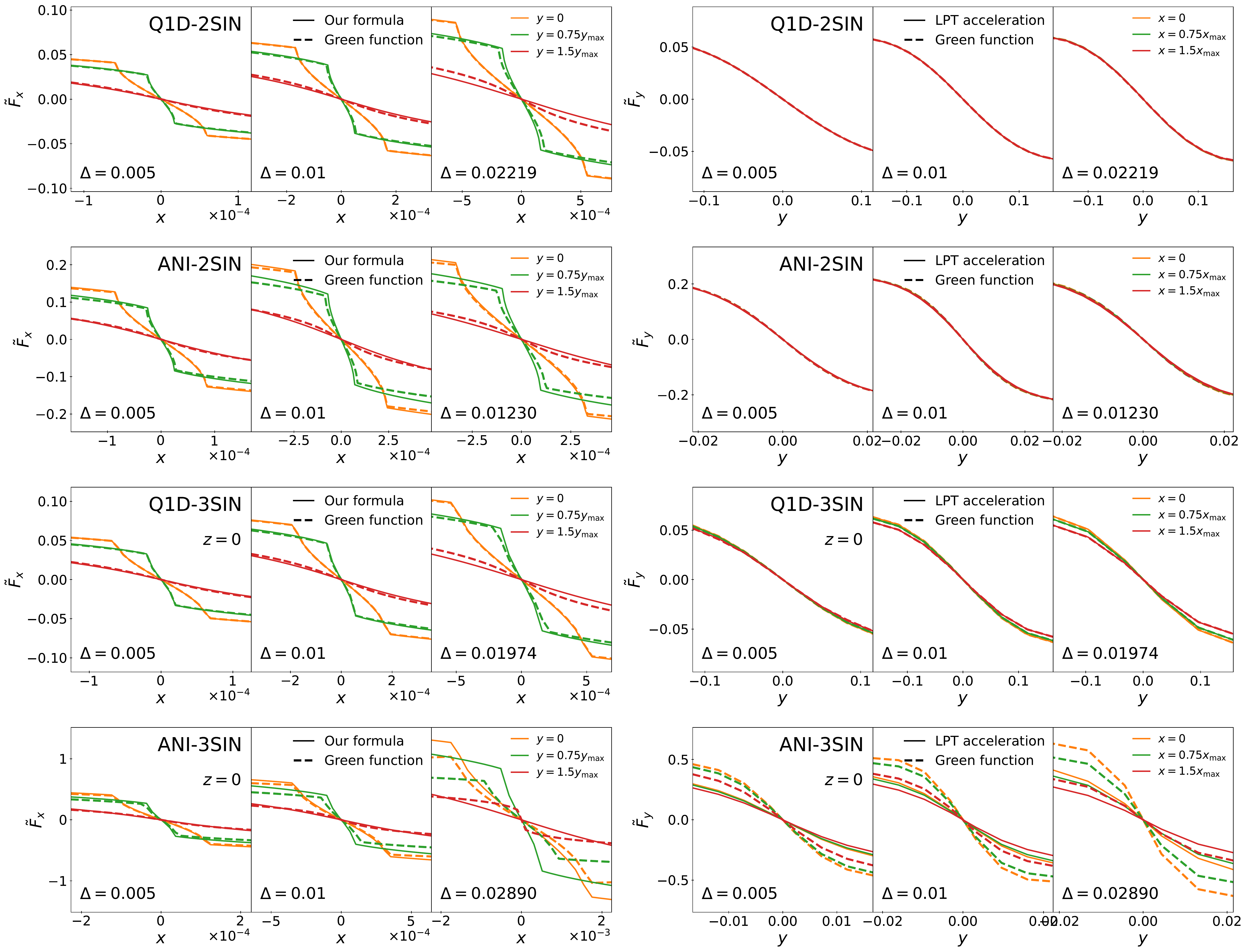}
\caption{
Time dependence of the $x$- and $y$-components of the force, for various sine wave initial conditions considered in this article, as indicated in the panels. On each row, the output time, traced by the parameter $\Delta$ in Eq.~(\ref{eq:deltadef}), increases from left to right, as indicated on each panel, with $\Delta=\Delta_{\rm sim}\,(a_{\rm sc}/a_{\rm sc}^{\rm LPT})$ on the right panels, where $\Delta_{\rm sim}$ corresponds to the output time in the simulations discussed in Sect.~\ref{sec: comparison} (see also Table~\ref{tab: initial conditions}). Note that in this last case, the value indicated on the figure is not $\Delta$ but $\Delta_{\rm sim}$, but the differences are small. The exact solution given by Eq.~(\ref{eq: Newton}) applied to the LPT displacement (dashes)  is compared to the analytical formula (\ref{eq: our formula}) for $F_{x}$ and with the LPT acceleration (\ref{eq: LPT acc multi}) for $F_{y}$ (solid lines). The two top and two bottom rows use 40LPT and 15LPT respectively to compute the displacement field and its acceleration. In the to bottom rows, we assume $z=0$.}
\label{fig: F time}
\end{figure*}
In the analyses performed in the two previous sections, we made sure that the assumptions of Sect.~\ref{sec: setup problem} were valid by taking a very small value of $\Delta=0.001$ in Eq.~(\ref{eq:deltadef}). In Fig.~\ref{fig: F time}, we now investigate how the results change when $\Delta$ increases, up to the values $\Delta_{\rm sim}^{\rm LPT}=\Delta_{\rm sim}\,(a_{\rm sc}/a_{\rm sc}^{\rm LPT})$ with $\Delta_{\rm sim}$ listed in Table~\ref{tab: initial conditions}, that are used for measurements of Sect.~\ref{sec: comparison} in {\tt ColDICE} simulations. 

We first focus on $F_{x}$ by examining left panels of Fig.~\ref{fig: F time}. As expected, the analytical prediction (\ref{eq: our formula}) deviates more and more from the exact solution as $\Delta$ increases. For a given value of $\Delta$, e.g. $\Delta=0.01$, deviations are slightly larger for ANI cases than Q1D cases, in agreement with intuition. Yet, even for non Q1D initial conditions, the local quasi one-dimensional nature of the dynamics dominates if $\Delta$ is small enough. Obviously our approach has its limits, as illustrated by bottom right panel of the left part of Fig.~\ref{fig: F time}. Importantly, deviations between dashes and solid lines on left part of Fig.~\ref{fig: F time} are not related to performances of high order LPT but instead, to our truncation to third order of the Taylor expansion in Eqs~(\ref{eq: Psi x expansion})--(\ref{eq: Psi z expansion}), as illustrated by Fig.~\ref{fig: x_qx} in Appendix~\ref{app: x_qx}. This also explains why deviations also increase with the value of $y$, when passing from the green to the red curve on left panels of Fig.~\ref{fig: F time}. We note that it would be possible to correct the solutions (\ref{eq: Cardano}) of three-value problem by including perturbatively higher order terms in the Taylor expansion in $\bm{q}$, that would most certainly improve the agreement between the solid and the dashed curve in the ANI-3SIN case, even for $\Delta=0.0289$. Of course, such a procedure cannot apply to arbitrarily large $\Delta$. It would moreover become pointless since LPT becomes a worse prescription of the dynamics as $\Delta$ increases. Indeed LPT has a finite convergence radius in time and does not take into account the feedback of the gravitational force inside the caustics, which is the objective we have in mind when computing the force field. 

Next, we focus on $F_y$ and examine right panels of Fig.~\ref{fig: F time},\footnote{The $z$-coordinate of the force, not shown here, would give very similar results.} which confirm what we found in previous section for $\Delta=0.001$: except for the ANI-3SIN case, the LPT acceleration given by Eq.~(\ref{eq: LPT acc multi}) agrees well with the exact solution (\ref{eq: Newton}) at all times considered. It stays pretty insensitive to the existence of the multi-stream region, as long as the pancake remains very thin and preserves the one dimensional nature of the local dynamics, which facilitates convergence of the LPT series. This is obviously not the case of ANI-3SIN, where $F_y$ presents significant variations with $x$ that increase with time, especially for the exact solution. Note that the good agreement between theory and exact solution is obviously greatly facilitated by 40LPT in 2D, while only 15LPT is used in the 3D cases. Still, although the Q1D-3SIN case uses only the 15LPT solution for the acceleration, theory and exact solutions still agree very well with each other. On the other hand, in the ANI-3SIN case, the LPT acceleration deviates significantly from the exact force given by Eq.~(\ref{eq: Newton}) and this even before shell-crossing, while the 15LPT displacement itself remains a very good approximation of the true displacement measured in simulations, even for $\Delta = \Delta_{\rm sim}^{\rm LPT}$, as shown below in Sect.~\ref{sec: comparison}.  
\section{Comparison to simulations}
\label{sec: comparison}
In the previous sections, we tested the accuracy of force field calculations by using LPT as a proxy of the dynamics. We now rely on actual measurements in Vlasov-Poisson simulations to test LPT itself, since it is used for computing the coefficients in Eqs.~(\ref{eq: xq 3})--(\ref{eq: zq 3}) that lead to the asymptotic expression (\ref{eq: our formula}) as well as the gravitational acceleration as a second time derivative of the LPT displacement in Eqs.~(\ref{eq:acc1D}) and (\ref{eq: LPT acc multi}).

The simulations we use to conduct the analyses were performed by \citet[][]{2021A&A...647A..66C} and STC with the public Vlasov code {\tt ColDICE}~\citep{2016JCoPh.321..644S}. This solver follows directly the evolution of a self-gravitating three-dimensional (two-dimensional) phase-space sheet in six-dimensional  (four-dimensional) phase-space with an adaptive tessellation of tetrahedra (triangles). The initial configuration uses a regular pattern of $n_{\rm s}$ vertices with null velocities to construct the tessellation. At the beginning of the simulations, this pattern is perturbed with Zel'dovich motion following Eq.~(\ref{eq:init_psi}).  During runtime, Poisson equation is solved in {\tt ColDICE} on a mesh of fixed resolution $n_{\rm g}$. Table~\ref{tab: initial conditions} indicates the values of $n_{\rm s}$ and $n_{\rm g}$ adopted for our runs, as well as the expansion factor $a_{\rm sim}$ of the snapshot used for the analyses slightly beyond first shell-crossing. Other technical details on the simulations we use are already provided in \citet[][]{2021A&A...647A..66C} and STC, so we do not repeat them here. In Appendix~\ref{app: force sim}, we explain how we measure the force field from the tessellation.

To compare LPT predictions to measurements in simulations, we proceed as in Sect.~\ref{sec: analytical ex} and synchronize high-order LPT solutions (here, 40LPT for the 2D case and 15LPT for the 3D case) as follows. First, LPT solutions are evolved to their own shell crossing time, designed in Table~\ref{tab: initial conditions} by $a^{\rm LPT}_{\rm sc}$ (namely $a^{(40)}_{\rm sc}$ and $a^{(15)}_{\rm sc}$ for 2D and 3D respectively, in the notations of Sect.~\ref{sec: analytical ex}). Next, time is advanced up to expansion factor $a = a^{\rm LPT}_{\rm sc} + (a_{\rm sim} - a_{\rm sc})$, where $a_{\rm sc}$ and  $a_{\rm sim}$ are respectively the shell-crossing time measured by STC in the simulations and the expansion factor of the snapshot we consider for the analyses, as listed in Table~\ref{tab: initial conditions}.  Note that the output times of the simulations, which are different for each initial condition, are, in terms of relative expansion factor values, closer to the shell crossing time in the following order: ANI-2SIN, Q1D-3SIN, Q1D-2SIN and ANI-3SIN (see the last column in Table~\ref{tab: initial conditions}).

\begin{figure*}
\centering
\includegraphics[width=\textwidth]{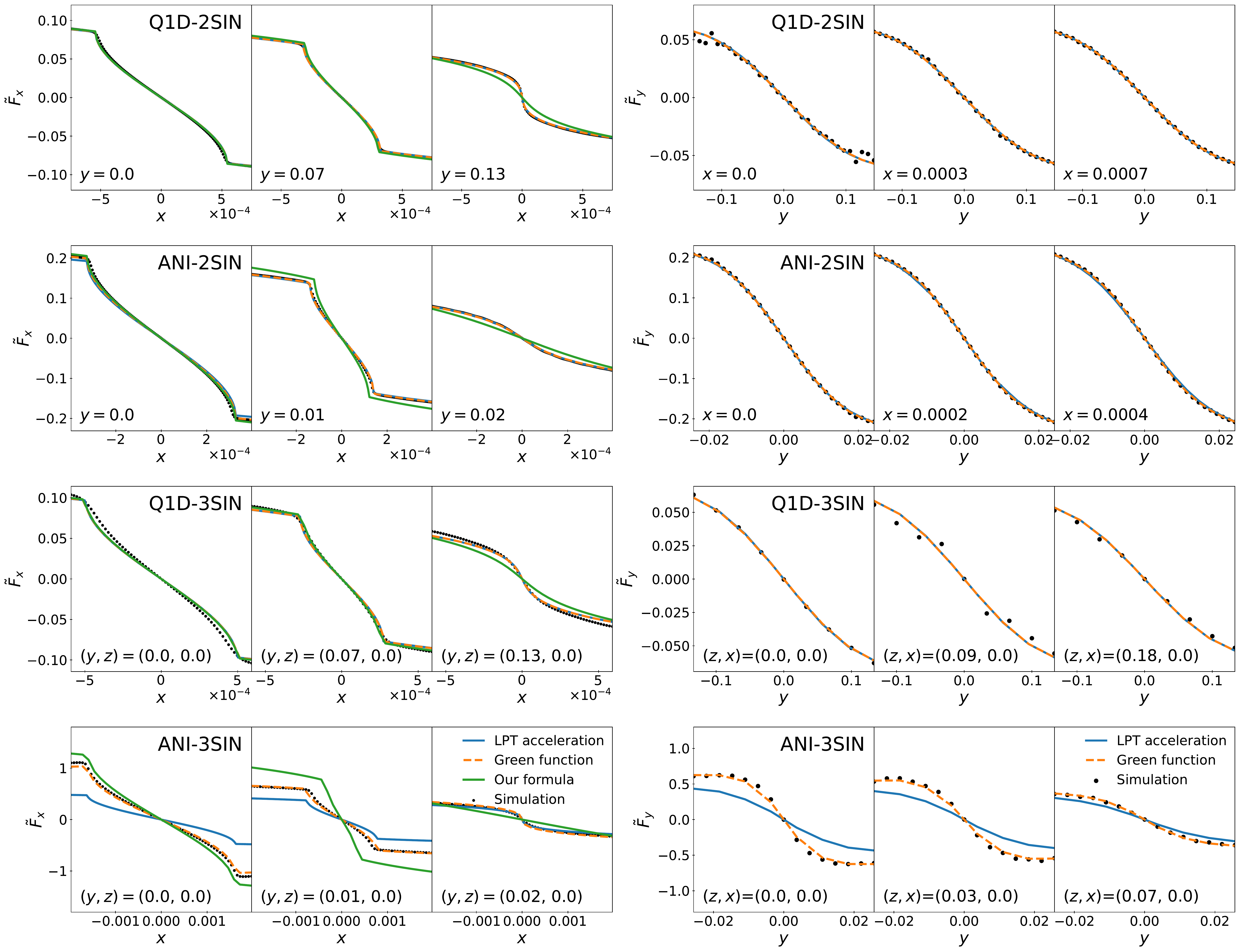}
\caption{Comparison between simulations and analytical predictions for $F_{x}$ (left panels) and $F_{y}$ (right panels) slightly beyond first shell-crossing time. From top to bottom, we consider Q1D-2SIN, ANI-2SIN, Q1D-3SIN, and ANI-3SIN initial conditions, respectively. For each initial condition, we present the force for three different values of $y$ and/or $z$ as indicated on each panel. The largest values of $y$ and $z$ are outside the multistream region. The dots stand for the measurements in the {\tt ColDICE} simulations. The green solid curves on left panels give the theoretical prediction for $F_x$ from the analytical formula (\ref{eq: our formula}). The solid blue lines correspond to the theoretical predictions for $F_{x}$ and $F_{y}$ obtained from the LPT acceleration using Eq.~(\ref{eq:acc1D}) (only in the multi-stream region) and Eq.~(\ref{eq: LPT acc multi}), respectively. The orange dashed lines represent the force field resulting from solving Poisson equation~(\ref{eq: Newton}) when using the LPT displacement instead of the supposedly exact positions of particles in the {\tt ColDICE} runs. 
}
\label{fig: sim}
\end{figure*}
Fig.~\ref{fig: sim} compares analytical predictions to measurements in the simulations of the force field for the various sine wave initial conditions introduced in Sect.~\ref{sec:sine_ini}, except the axisymmetric ones which are discussed further below. The first thing to notice when examining this figure is the excellent agreement between the measurements (black dots) and the numerical solution of Poisson equation based on the Green function approach, Eq.~(\ref{eq: Newton}), when applied to the density field sourced by the LPT displacement after synchronisation (orange dashes). This confirms the earlier investigations of STC on the density field, which compared post-collapse predictions based on the ballistic approximation, where the velocity field is frozen at collapse time, to the same simulations data. At the times considered here, the ballistic approximation should not differ much from LPT predictions pushed beyond shell-crossing as performed in the present work. Such a good agreement between LPT and simulations fully validates the conclusions of the analyses of Sect.~\ref{sec: analytical ex}. Thus, in Fig.~\ref{fig: sim}, we observe the same matches/discrepancies between the solid green (left panels)/blue (right panels) curves and the black dots/orange dashes that we can discern in Fig.~\ref{fig: F time}, between the solid curves and the dashes. This confirms again the validity of our theoretical predictions when two conditions are met: (i) sufficiently short period of time after collapse, so that our local Taylor expansion of the displacement field remains accurate enough to estimate the component of the force along the shell-crossing direction (orthogonal to the pancake) with Eq.~(\ref{eq: our formula}), and (ii) sufficiently high LPT order for the LPT displacement to be accurate around shell-crossing time, as well as its second time derivative used in Eq.~(\ref{eq: LPT acc multi}) as an approximation for the component of the force orthogonal to the direction of collapse (coplanar with the pancake). Obviously, these two conditions are not met for ANI-3SIN as already discussed in detail in Sect.~\ref{sec: analytical ex}, but are facilitated in the Q1D cases. 

Interestingly, Eq.~(\ref{eq:acc1D}) (solid blue curves on left panels of Fig.~\ref{fig: sim}) approximates very well the $x$-component of the force, except again for ANI-3SIN. Summing up correctly the LPT accelerations in the multi-stream regions using either Eq.~(\ref{eq:acc1D}) or Eq.~(\ref{eq: LPT acc multi}) also provides, shortly after shell-crossing, a very good self-consistent approximation of all the components of the force field provided that the LPT series is converged.

\begin{figure}
\centering
\includegraphics[width=\columnwidth]{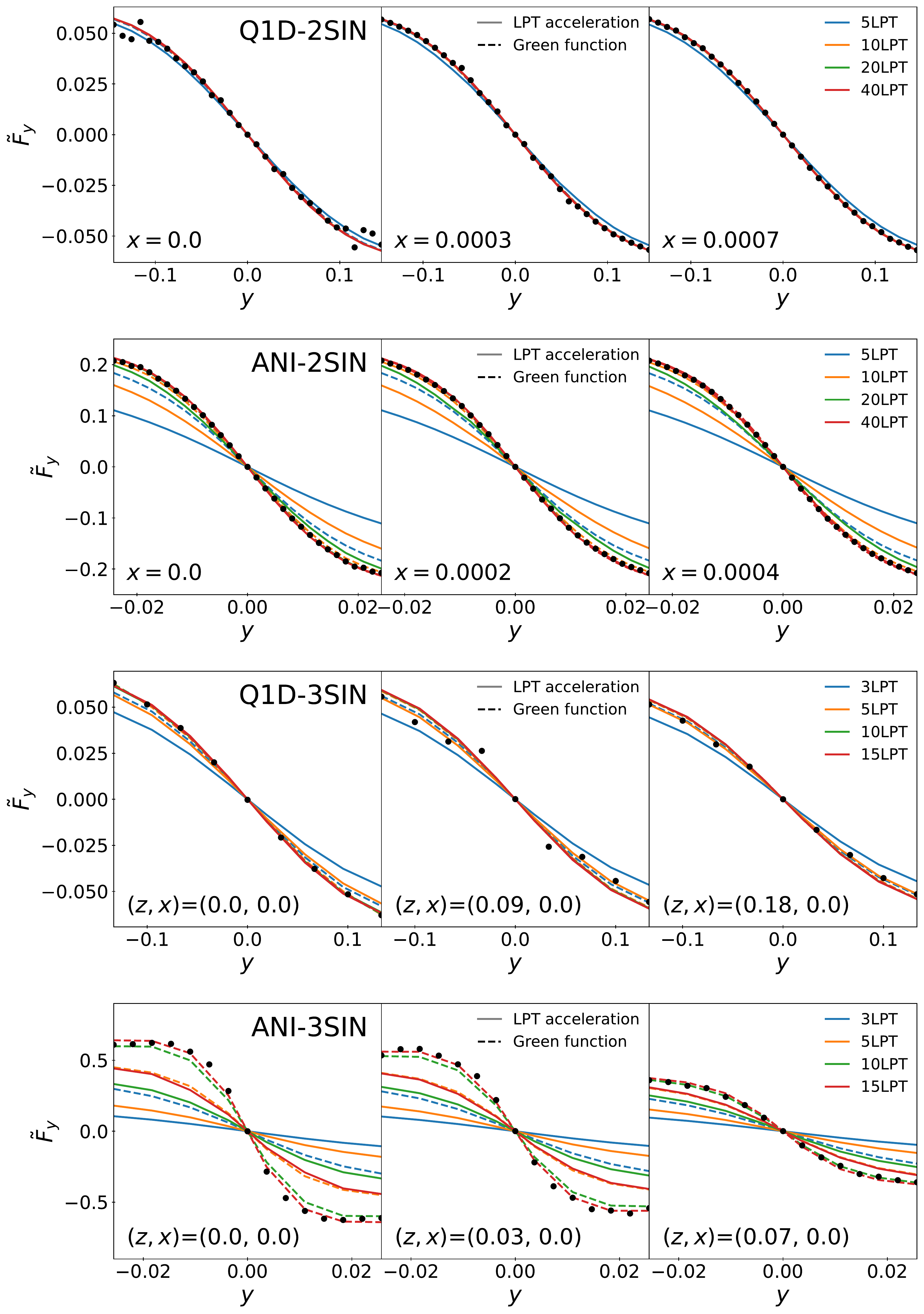}
\caption{Same as in the right panels of Fig.~\ref{fig: sim}, but without synchronisation to collapse time at each LPT order, i.e., for a given initial condition, all the curves are calculated at the same expansion factor, $a_{\rm sim}$. In addition, several LPT orders are considered as indicated on right panels, to illustrate the convergence speed.}
\label{fig: sim wo sync}
\end{figure}
The synchronisation process plays an important role in the analytical prediction of the force along the direction of shell-crossing, $F_x$, because collapse time changes with LPT order and the effect of shell-crossing is dramatic on $F_x$. However, this is not really the case for the orthogonal component --coplanar with the pancake-- that preserves pretty much the single-stream behaviour of the motion, as already discussed in Sect.~\ref{sec: analytical study Fy}. We double check again this property by reproducing the right panels of Fig.~\ref{fig: sim} on Fig.~\ref{fig: sim wo sync} but without synchronisation of LPT predictions. For the $40^{\rm th}$ LPT order and the $15^{\rm th}$ LPT order respectively in 2D and 3D, we indeed distinguish no difference between the two figures. Additionally, various LPT orders are considered on Fig.~\ref{fig: sim wo sync}. Interestingly, the LPT accelerations (solid lines) show slow convergence, while the forces computed with the Green function method from the LPT displacements (Eq.~\ref{eq: Newton}, dashed lines) show very fast convergence, even for ANI-3SIN. In other words, as mentioned in Sect.~\ref{sec: analytical study Fy}, Eq.~(\ref{eq: Newton}) represents a {\em resummation} of LPT that provides a much more accurate description of the gravitational acceleration than the second time derivative of the LPT displacement. This property might turn very useful for future applications.

\begin{figure}
\centering
\includegraphics[width=\columnwidth]{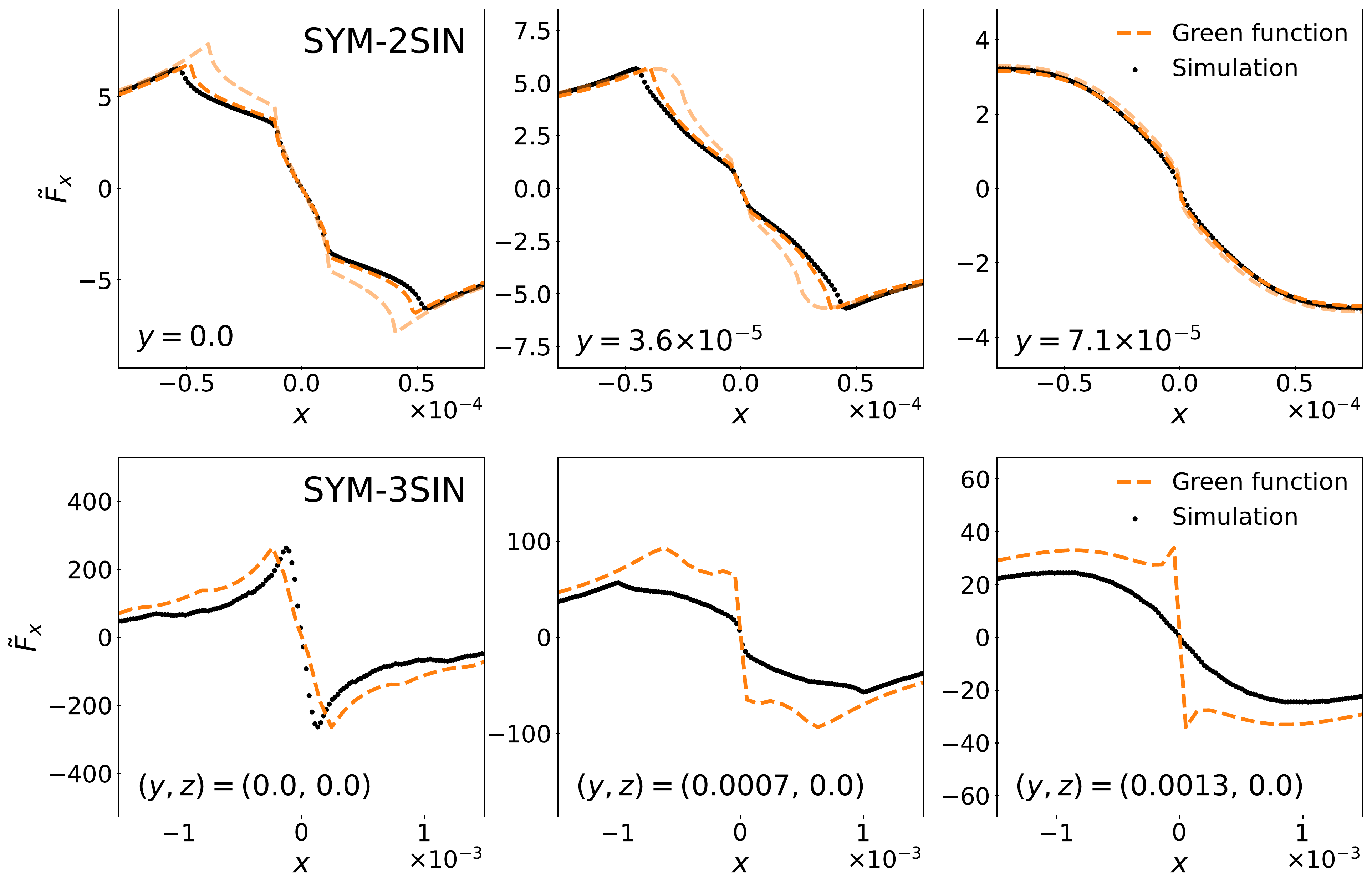}
\caption{
Same as Fig.~\ref{fig: sim} but for the SYM-2SIN (top panels) and SYM-3SIN (bottom panels) initial conditions, except that they are no blue nor green curves, because our analytical recipes apply only to shell-crossing occurring along one direction.  Only the $x$-component of the force is shown in Fig.~\ref{fig: sim SYM} due to the symmetric nature of the system. In addition, the 15LPT prediction from Eq.~(\ref{eq: Newton}) is shown as a light orange dashed curve in the 2D case.  }
\label{fig: sim SYM}
\end{figure}

Before closing this section, we examine in Fig.~\ref{fig: sim SYM} axial-symmetric initial conditions, SYM-2SIN and SYM-3SIN. In this case, shell-crossing occurs simultaneously along all the axes of the dynamics, so we expect qualitatively different behaviour from the Q1D and ANI cases. Strictly speaking axial-symmetric configurations have zero weight from the statistical point of view, but can in practice still be present at the coarse level, e.g. very high peak in random Gaussian fields which are expected to be rounder \citep[see, e.g.,][]{1986ApJ...304...15B}. 

The axisymmetric case does not correspond to the superposition of 2 and 3 pancakes in the 2D and 3D cases, respectively \citep[see][for analytical predictions of the gravitational potential under these assumptions]{Gurevich_1995}. Indeed, the caustic structure stemming from simultaneous shell-crossings along several directions is convoluted, as shown by STC, which translates into multiple discontinuities of the derivative of the force field along each axis. For instance, on top left panel of Fig.~\ref{fig: sim SYM}, there are four sharp transition points on the force field instead of two on top left panel of Fig.~\ref{fig: sim}. Even if the analytical predictions discussed in Sect.~\ref{sec: formula Fx} are irrelevant for SYM cases, it is still possible to use the Green function approach in Eq.~(\ref{eq: Newton}) with the high order LPT displacement field used as a proxy for the dynamics, as indicated by the dashed curves on Fig.~\ref{fig: sim SYM}. While 40LPT successfully reproduces the post-collapse gravitational field in the 2D case, 15LPT is not accurate enough. Turning to the 3D case, 15LPT is not even good from the qualitative point of view, as already noticed for STC for the density field itself. However, SYM-3SIN is further away from shell-crossing than SYM-2SIN in terms of the parameter $\Delta$ as shown in last column of Table~\ref{tab: initial conditions}, which explains partly the worse performance of LPT predictions for SYM-3SIN compared to SYM-2SIN. Another and more obvious reason for this lies in the much higher contrasts expected in SYM-3SIN compared to SYM-2SIN, which can introduce non negligible feedback effects from the multistream region, even very shortly after shell-crossing, as already extensively discussed in STC.

\section{Summary}
\label{sec: summary}
In this article, we studied the gravitational force field generated by pancakes shortly after collapse. Restricting to the case where the displacement field sourcing the pancake is locally symmetric, we derived approximations for the force field combining fundamentals of catastrophe theory and high order Lagrangian perturbation theory (LPT) that we carefully validated on systems seeded by two or three sine waves initial conditions. Our analyses included comparisons of theoretical predictions to measurements in Vlasov simulations performed with the public code {\tt ColDICE}. The important assumption is that the time lapse after collapse, quantified here by the parameter $\Delta$ given in Eq.~(\ref{eq:deltadef}), is very short, allowing one to assume that the multistream region is locally infinitely thin. The main results of our work can be summarized as follows:
\begin{enumerate}
\item[(i)] The calculation of the component $F_x$ of the gravitational force aligned with the direction of shell-crossing (that is, orthogonal to the pancake) comes down to solving a three value problem that reduces to the resolution of a third order polynomial in the limit $\Delta \ll 1$. This process is very analogous, not surprisingly, to the 1D case, and leads to explicit expressions in terms of Eulerian coordinates, Eqs.~(\ref{eq:qyqx}), (\ref{eq:Aconst}), (\ref{eq:Bconst}), (\ref{eq: Cardano}) and (\ref{eq: our formula}). Calculation of the various time dependent coefficients intervening in the expression of $F_x$ rely on a Taylor expansion of the LPT displacement at third order in the Lagrangian position, while staying within the convergence radius of the displacement expanded as a high order series in time. 
  
\item[(ii)] The component $F_y$ (and $F_z$) of the gravitational force orthogonal to the direction of shell-crossing (that is, in the same plane as the pancake) is rather insensitive to the presence of caustics. It can then be predicted with the usual LPT acceleration, that is the second time derivative of the LPT displacement, as long as the LPT acceleration converges as a series expansion in time. While Eq.~(\ref{eq: LPT acc multi}) can be used to perform averaging over several streams, it is not absolutely needed. However the three value problem remains to be solved, either numerically as we did in this work, or approximately using Eqs.~(\ref{eq:qyqx}),  (\ref{eq:Aconst}), (\ref{eq:Bconst}) and (\ref{eq: Cardano}). 

\item[(iii)] We noted that Eq.~(\ref{eq: our formula}) is asymptotically equivalent to Eq.~(\ref{eq:acc1D}) with the $x$-component of the acceleration given by the second time derivative of the LPT displacement, which in turn makes the approach fully consistent with (ii).

\item[(iv)] Much higher LPT order is needed for the acceleration to converge than for the displacement, which is a natural consequence of the properties of the LPT series, of which the convergence speed decreases with successive time derivatives~\citep[see e.g.,][]{2023arXiv230312832R}. However, one can solve numerically Poisson equation using the Green function approach embodied by integral Eq.~(\ref{eq: Newton}) based on the knowledge of the LPT displacement to reach --as obvious-- a convergence level as good as for the displacement. Hence Eq.~(\ref{eq: Newton}) acts as a resummation procedure of LPT to compute accurately the gravitational field.

\item[(v)] While quasi one-dimensional initial conditions facilitate convergence of perturbation theory predictions, our calculations apply also to pancakes seeded by peaks with a general ellipsoid shape and remain accurate as long as the pancake remains very flat, but become wrong in the extreme case where collapse takes place simultaneously along all the axes of the dynamics. 
\end{enumerate}
In this work, we have used LPT as an approximation of the dynamics shortly beyond collapse, but there are limitations to such an approach due to the finite radius of convergence with respect to time of the perturbative series. Another safer point of view, adopted and tested successfully by STC on our sine wave cases, consists in using a ballistic approximation, where velocities (and in the present case LPT accelerations) are frozen from collapse. While we did not adopt it here for simplicity, it would be easy to implement this ballistic approximation to compute the gravitational force field of pancakes, instead of the procedure described in points (i) and (ii) above. The advantage of the ballistic approach is to overcome the problem of LPT convergence which is known to be guaranteed in practice up to collapse~\citep[see, e.g.,][STC]{2021MNRAS.501L..71R}. 
 
The calculations presented in this work can in principle be generalised to an arbitrary (smooth and non degenerate) displacement field. Clearly, the case of a displacement field deriving from a potential in Lagrangian space (i.e. vorticity free in Lagrangian space) can be treated easily. While not fully representative of the exact dynamics since zero vorticity prior to shell-crossing is only expected in Eulerian space, a vorticity free displacement in Lagrangian space remains a very good approximation up to second order in the LPT framework, that includes Zel'dovich approximation. In the case of a displacement field deriving from a potential, it is easy to realize with the proper combination of affine transformations both in Lagrangian and Eulerian space, that one can obtain equations similar to Eqs.~(\ref{eq: xq 3})--(\ref{eq: zq 3}), but with additional $q_x^2 q_y$, $q_x^2 q_z$, $q_y^3$ and $q_z^3$ terms in the right member of Eq.~(\ref{eq: xq 3}), which makes solving the three-value problem and the force field calculation slightly more involved. We postpone more general analyses which do not impose the local symmetries (\ref{eq:sym1})--(\ref{eq:sym3}) to future work.

The calculation of the force field generated in the vicinity of a protopancake represents the first step for an accurate treatment of the dynamics in the multi-stream regime, by computing corrections to the motion inside the pancakes due to the force backreaction from the multistream regions, extending thereby the 1D calculations of~\citet{2015MNRAS.446.2902C,2017MNRAS.470.4858T,2021MNRAS.505L..90R} to the 3D case. Of course, this approach remains limited, as it is expected to work only shortly after shell-crossing, even though it might be possible to combine it with an adaptive smoothing procedure to accurately predict large-scale structure statistics such as the power spectrum, even in the nonlinear regime \citep[see, e.g.,][for the 1D case]{2017MNRAS.470.4858T,2020MNRAS.499.1769H}. Shell-crossing can also locally take place along other axes of the dynamics, leading to the formation of protofilaments and protohaloes. This is also followed by violent relaxation, that is quick folding of the phase-space sheet in multiple directions. Understanding in details the early evolution of protopancakes remains crucial to understand how multiflow dynamics is initiated and how this affects the early evolution of the statistics of the large scale matter distribution. 

\section*{Acknowledgements}

We thank C.~Rampf for reading and commenting on the draft.
SS is supported by JSPS Overseas Research Fellowships.
This work was supported in part by MEXT/JSPS KAKENHI Grant Numbers JP20H05861 and JP21H01081 (AT), ANR grant ANR-13-MONU-0003 (SC), as well as Programme National Cosmology et Galaxies (PNCG) of CNRS/INSU with INP and IN2P3, co-funded by CEA and CNES (SC).
Numerical computation with {\tt ColDICE} was carried out using the HPC resources of CINES (Occigen supercomputer) under the GENCI allocations c2016047568, 2017-A0040407568 and 2018-A0040407568. Post-treatment of {\tt ColDICE} data were performed on HORIZON cluster of Institut d'Astrophysique de Paris.

\bibliography{ref}
\bibliographystyle{aa}

\begin{appendix}
\section{Series expansion of the displacement field}
\label{app: x_qx}
\begin{figure}
\centering
\includegraphics[width=\columnwidth]{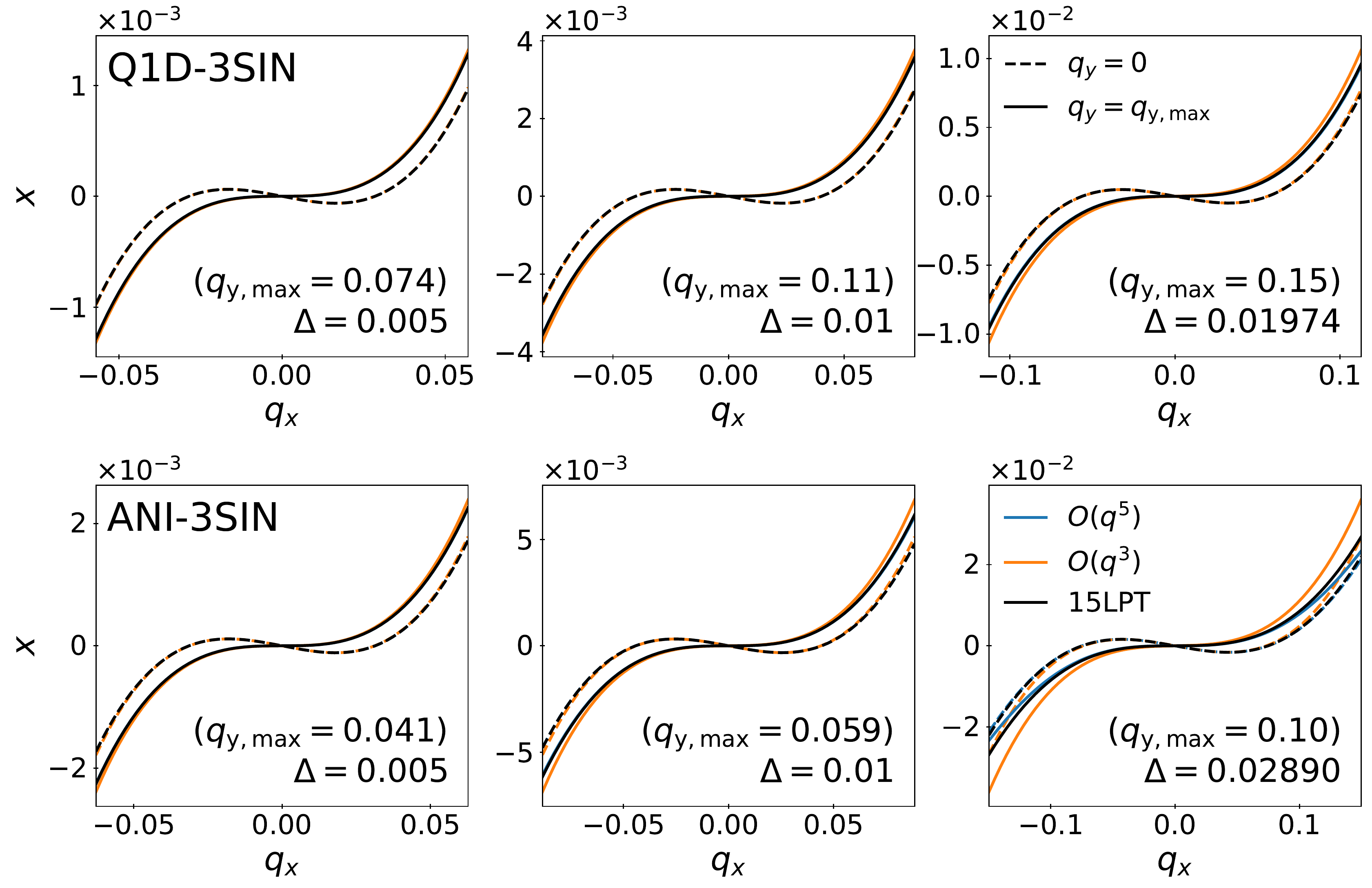}
\caption{$x$ coordinate as a function of $q_x$ for Q1D-3SIN (top panels) and ANI-3SIN (bottom panels) initial conditions at various output times, with $\Delta=0.005$ (left), $0.01$ (centre), and $\Delta_{\rm sim}=a_{\rm sim}/a_{\rm sc}-1$ as given in right column Table~\ref{tab: initial conditions} (right). The solid and dashed curves correspond to $q_y=0$ and $q_y\simeq q_y^{\rm max}$, where $q_y^{\rm max}$ is the the max extension of the caustic in Lagrangian space along $q_y$ axis, while $q_z=0$. 
We present the results predicted by the pure 15LPT solution (black) and its Taylor expansion expression in the Lagrangian coordinate up to up to $\mathcal{O}(q^{3})$ (orange) and up to $\mathcal{O}(q^{5})$ (blue). 
}
\label{fig: x_qx}
\end{figure}

In Sect.~\ref{sec: setup problem}, we derived an analytical expression for computing the $x$-component of gravitational force, starting from the Taylor expansion of the displacement field with respect to the Lagrangian coordinate up to third order in $\bm{q}$, given in Eqs.~(\ref{eq: xq 1})--(\ref{eq: zq 1}). Fig.~\ref{fig: x_qx} tests the accuracy of Eqs.~(\ref{eq: xq 1})--(\ref{eq: zq 1}) in $q_{x}$-$x$ space for our non-axisymmetric three-sine wave setups.  Clearly, at the dynamical times considered in this paper, a third order approximation for the displacement field is sufficient in the Q1D case, but significant deviations can be noticed on bottom right panel of the figure. As expected, these deviations increase when moving away from the origin of coordinates, hence with increasing, $|x|$, $|y|$ (and $|z|$ in 3D). They explain, at least partly, the discrepancies between theory and exact solution in Fig.~\ref{fig: F time}. Another significant source of error is the approximation of $y$ and $z$ coordinates at linear order in $\bm{q}$ (Eqs.~\ref{eq: yq 3} and \ref{eq: zq 3}), that we do not examine here. Note that, as long as higher order corrections remain small, it would be possible to correct perturbatively the three-value solution derived from the third order Taylor expansion to include higher order contributions. Here, we notice that $5^{\rm th}$ order already brings very significant improvements, even if we should also check what happens in other planes.

\section{Calculation of the force: truncation of integral (\ref{eq: Newton})}
\label{app: different G}
Throughout this article, we compare our analytical predictions for the force to direct calculations of the integral~(\ref{eq: Newton}) over a finite interval, $[-q_{\rm max}, q_{\rm max}]$. Our systems seeded by sine wave initial conditions are periodic, hence the problem comes down to compute the gravitational contribution of each matter element inside the simulation volume of size $L$, plus all the periodic replica. Many techniques exist in the literature to perform such a calculation quickly, for instance Ewald summation \citep[see, e.g.][]{1991ApJS...75..231H}. Our brute force technique allows us to have rather accurate estimates of pairwise interactions between closeby elements of mass, which is necessary given the very thin nature of the caustic structures, but fails to account for the contribution of all remote elements, including replica, which can make the force field calculation inaccurate.

\begin{figure*}
\centering
\includegraphics[width=0.9\textwidth]{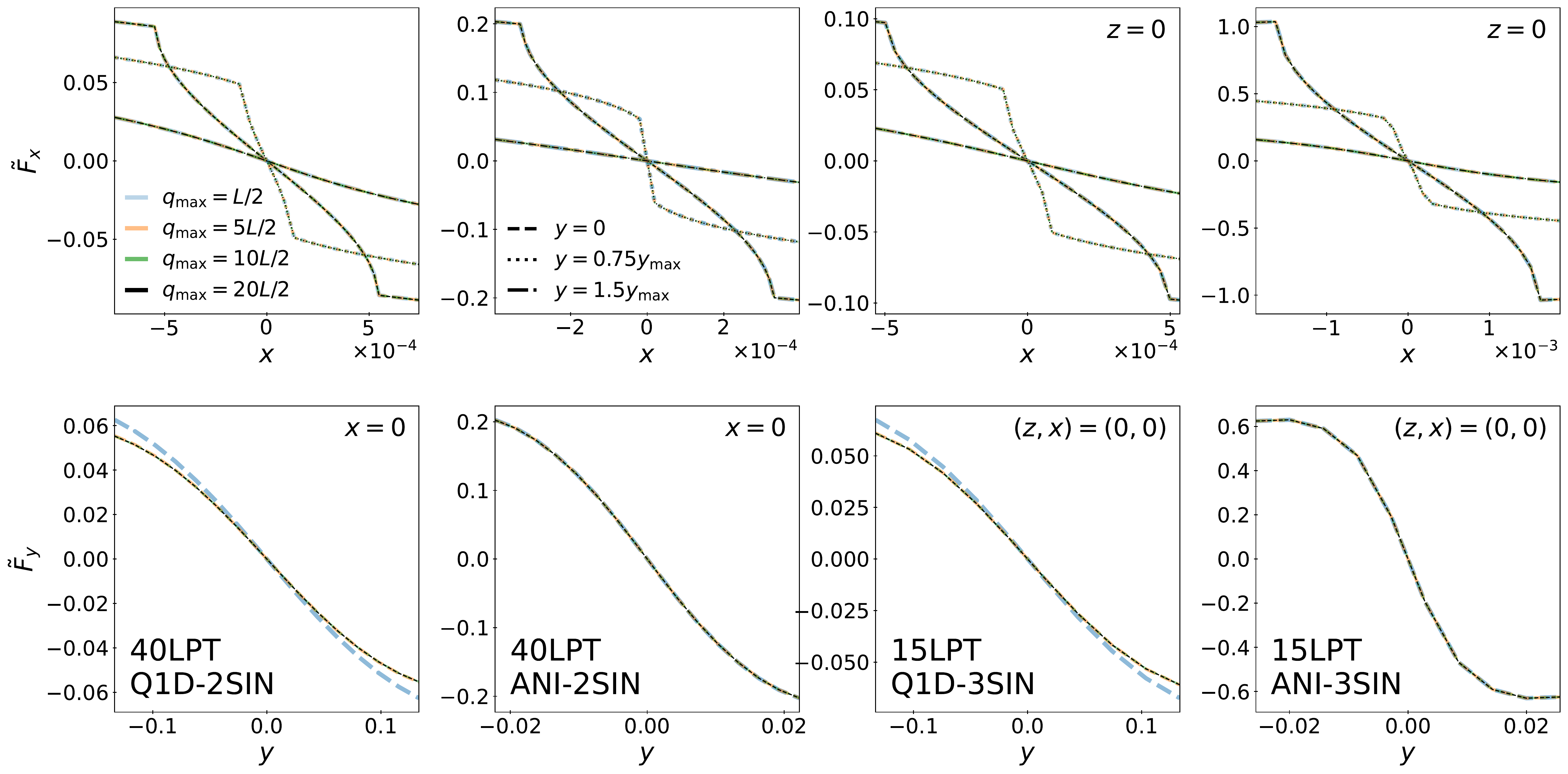}
\caption{Tests of the choice of the bounds of integral (\ref{eq: Newton}). The $x$ and $y$ coordinates of the force are plotted, respectively, on top and bottom panels, for various choices of the integration range $[-q_{\rm max},q_{\rm max}]$ with $q_{\rm max}$ ranging form $L/2$ to $20L/2$ as indicated on upper left panel. From left to right, we consider Q1D-2SIN, ANI-2SIN, Q1D-3SIN, and ANI-3SIN initial conditions. On top panels, various values of $y$ are considered as indicated on second panel of the top row, while the bottom panels only assume $x=0$ since other values of $x$ would not differ significantly. We also set $z=0$ in the three-dimensional cases considered in two top-right and tow bottom-right panels. Note that the output time chosen in this figure is the same as in Fig.~\ref{fig: sim}. We see that all the curves corresponding to different values of $q_{\rm max}$ are superimposed on each other, except for $q_{\rm max}=L/2$ for the Q1D cases.}
\label{fig: replica}
\end{figure*}

To validate the choice of our somewhat simplistic technique, Fig.~\ref{fig: replica} presents the $x$- (top panels) and $y$-components (bottom panels) of the force for our (non axisymmetric) sine wave initial conditions, restricting to the local neighbourhood of the pancake. All the output times are the same as in Fig.~\ref{fig: sim}, which compares analytical predictions to simulations. We see that the $x$-component of the force at the scales of interest is dominated by the influence of the caustics and is thus quite insensitive to the replica given these small values of $x$, so a truncation of the integral (\ref{eq: Newton}) at $q_{\rm max}=L/2$ is enough for all the initial conditions (including SYM-2SIN and SYM-3SIN, not shown in Fig.~\ref{fig: sim}). On the other hand, the $y$-component of the force can be significantly affected by the replicas for Q1D initial conditions, but from lower panels of Fig.~\ref{fig: sim}, $q_{\rm max}=5 L/2$ seems sufficient for adequate convergence of the integral (\ref{eq: Newton}). In practice, we adopted the values listed in Table~\ref{tab:param} for the measurements in the {\tt ColDICE} simulations, and $q_{\rm max}=L/2$  for $F_{x}$, $q_{\rm max}=20L/2$ for $F_{y,z}$, for other calculations using the LPT displacement in Eq.~(\ref{eq: Newton}).

\section{Force measurement in simulations}
\label{app: force sim}
\begin{table*}[htp]
\centering
\begin{tabular}{ccccccc}
\hline
Designation   & $q_{\rm max}$ & $x_{\rm max}$ &  $y_{\rm max}$ & $z_{\rm max}$ & $\ell_{\rm max}$ & $\varepsilon$ \\
\hline
{\it Quasi 1D}\\
Q1D-2SIN & $L/2$ \ for\ $F_x$,\ $30.5 L$\ otherwise & $0.00075$ & $0.15$ &             & 2 & $1.1718750 \times 10^{-5}$ \\
Q1D-3SIN & $L/2$ \ for\ $F_x$,\ $5 L/2$\ otherwise   & $0.0006$   & $0.15$ & $0.20$ & 4 & $9.375  \times 10^{-6}$\\
\hline
{\it Anisotropic}\\
ANI-2SIN & $L/2$ \ for\ $F_x$,\ $30.5 L$\ otherwise & $0.0004$ & $0.025$    &              & 2 & $6.25\times 10^{-6}$ \\
  ANI-3SIN & $L/2$ \ for\ $F_x$,\ $ 7 L/2$\ otherwise   &  $0.002$  & $ 0.0275$ & $ 0.07$ &  2\ for \ $q_{\rm max}=L/2$ & $3.125 \times 10^{-5}$\\
  & & & & & \ 1 \ for\ $q_{\rm max}=7 L/2$ & \\
\hline
{\it Axial-symmetric}\\
SYM-2SIN & $L/2$  &  $0.00008$ & $ 0.00008$ &    &  2 & $1.25 \times 10^{-6}$ \\
SYM-3SIN & $L/2$  &  $ 0.0015$ & $ 0.0015$ & $0.0015$ & 0 & $2.34375 \times 10^{-5}$\\
\hline
\end{tabular}
\caption[]{Parameters intervening in the measurement of the force field in the Vlasov simulations as described in Appendix~\ref{app: force sim}. From left to right: designation of the run, truncation parameter $q_{\rm max}$, coordinates $x_{\rm max}$, $y_{\rm max}$ and $z_{\rm max}$ of the upper right corner of the rectangular region where higher sampling of simplices is perform using $\ell_{\rm max}$ successive refinements as indicated in the next column; finally, $\varepsilon$ is the softening parameter of the force field (Eq.~\ref{eq:forcesoft}).}
\label{tab:param}
\end{table*}

In {\tt ColDICE}, the dark matter distribution is represented with an adaptive tessellation of simplices, that is an ensemble of connected triangles and tetrahedra, respectively in two and three dimensions. To compute the force field $\tilde{\bm{F}}(\bm{x})$ at a given point of space, we employ a direct approach consisting in replacing each simplex with a set of particles. To do so, if needed, each simplex is refined isotropically $\ell_{\rm max}$ times. At the end, each sub-simplex is replaced with a single particle lying at the barycentre of the its vertices. For best accuracy, the refinement process exploits the quadratic nature of each simplex thanks to additional tracers used in {\tt ColDICE}. To avoid divergences due to the singular nature of the force field induced by a point particle distribution, we introduce additional softening as follows:
\begin{align}
  \tilde{\bm{F}}(\bm{x}) =\frac{m}{2^{d-2} \times 2 \pi} \frac{ \bm{x}-\bm{x}_0}{(|\bm{x}-\bm{x}_0|^2+\varepsilon^2)^{d/2}},
  \label{eq:forcesoft}
\end{align}
where, $d=2$ or 3 is the dimension of space, $m$ and $\bm{x}_0$ are respectively the (normalised) mass and the position of the particle. 
This equation does not account for the background correction proportional to $\bm{x}$ in equation (\ref{eq: Newton}), that has to be added at the end of the calculation. Depending on the coordinate of the force field considered, a set of periodic replica of each particle with positions
\begin{align}
  \bm{x}_0+(i,j,k)\, L,
  \label{eq:repli}
\end{align}
can contribute, with integers $i,j,k \in [-n_{\rm rep}, n_{\rm rep}]$, which is equivalent to restrict the integral (\ref{eq: Newton}) to the interval $[-q_{\rm max},q_{\rm max}]$, with $q_{\rm max}=(n_{\rm rep}+1/2) L$. The values of $q_{\rm max}$ we adopted are listed in Table~\ref{tab:param}. They are large enough according to the convergence tests discussed in Appendix~\ref{app: different G}. 

Because we examine the system just after shell-crossing, it is important to have an accurate description of the phase-space sheet inside a region containing the caustics. This region is chosen to be a thin rectangular parallellepiped covering the intervals $x \in [-x_{\rm max},x_{\rm max}]$, $y \in [-y_{\rm max},y_{\rm max}]$ and $z \in [-z_{\rm max},z_{\rm max}]$ (the latter in 3D only) in each dimension. Outside this region, each simplex is replaced with a single particle and inside it, each simplex is adaptively refined $\ell_{\rm max} \geq 1$ times, with a value of $\ell_{\rm max}$ depending on initial conditions.

The various parameters introduced in this appendix are listed in Table~\ref{tab:param}, which completes Table~\ref{tab: initial conditions} of the main text. Note for instance that $\varepsilon=x_{\rm max}/64$. Because of the high mass resolution of the 2D simulations ($n_{\rm s}=2048$), is was enough to take $\ell_{\rm max}=2$ for Q1D-2SIN, ANI-2SIN and SYM-2SIN. Furthermore, due to the lower cost of the force calculation in 2D, many periodic replica could be used, $n_{\rm rep}=30$. On the other hand, the 3D case is much more involved computationally, which imposes us to adopt a much smaller value of $n_{\rm rep}$, yet still large enough according to the tests performed in Appendix~\ref{app: different G}. The most delicate case was Q1D-3SIN, where $\ell_{\rm max}=4$ was necessary to have (barely) sufficiently accurate measurements of the force field. There are two reasons for this. First, in this simulation, the initial tessellation has a lower number of simplices, $6 \times n_{\rm s}^3$, with $n_{\rm s}=256$, compared to $n_{\rm s}=512$ for ANI-3SIN and SYM-3SIN, which already imposes at least one additional refinement level in Q1D-3SIN compared to the two other cases. Second, the discrete nature of the sampling of the phase-space sheet we adopted can introduce systematic biases on the force field, which are stronger in Q1D settings due to the fact that particles, that initially form a regular mesh, tend to cluster together much more along the $x$ direction inside and in the vicinity of the caustics than in orthogonal direction(s). This induces artificial fluctuations on the force field, especially on axes orthogonal to $x$ (see e.g. middle panel of the right group in third row of Fig.~\ref{fig: sim}), that would not appear with a proper, smooth representation of the phase-space sheet. While it would be possible (but not trivial) to implement a more optimal calculation of the force field, we did not find it necessary for our analyses.
\end{appendix}

\end{document}